# Prompt Engineering

For Students of Medicine and Their Teachers

Thomas F. Heston, MD

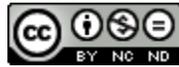

**Medical Disclaimer**

This book is intended to provide helpful and informative material on the subject matter covered. It is not meant to be used, nor should it be used, to diagnose or treat any medical condition. **For diagnosis or treatment of any medical problem, consult your own physician.**

The author and publisher are not responsible for any specific health needs that may require medical supervision and are not liable for any damages or negative consequences from any treatment, action, application, or preparation to any person reading or following the information in this book. References are provided for informational purposes only and do not constitute an endorsement of any websites or other sources.

**Please consult with your healthcare provider if you have any questions or concerns.** The information in this book is provided "as is" without warranty of any kind, express or implied, including but not limited to the warranties of merchantability, fitness for a particular purpose, and non-infringement. In no event shall the author or copyright owner be liable for any claim, damages, or other liability, whether in an action of contract, tort, or otherwise, arising from, out of, or in connection with the information provided in this book.

**This book represents the personal views and opinions of the author. It is for informational and entertainment purposes and does not constitute medical advice.**

**Funding Information:** Self-funded, no external funding.

**Conflict of Interests:** No competing interests.

**Ethical Approval:** This study did not involve human or animal research.

# Table of Contents

**Medical Disclaimer**

This book is intended to provide helpful and informative material on the subject matter covered. It is not meant to be used, nor should it be used, to diagnose or treat any medical condition. **For diagnosis or treatment of any medical problem, consult your own physician.**

The author and publisher are not responsible for any specific health needs that may require medical supervision and are not liable for any damages or negative consequences from any treatment, action, application, or preparation to any person reading or following the information in this book. References are provided for informational purposes only and do not constitute an endorsement of any websites or other sources.

**Please consult with your healthcare provider if you have any questions or concerns.** The information in this book is provided "as is" without warranty of any kind, express or implied, including but not limited to the warranties of merchantability, fitness for a particular purpose, and non-infringement. In no event shall the author or copyright owner be liable for any claim, damages, or other liability, whether in an action of contract, tort, or otherwise, arising from, out of, or in connection with the information provided in this book.

**This book represents the personal views and opinions of the author. It is for informational and entertainment purposes and does not constitute medical advice.**

**Prompt Engineering For Students of Medicine and Their Teachers**

**Author:** Thomas F. Heston, MD

**Publisher:** MegaSimple.com LLC, Las Vegas, NV

**Copyright** updated 8/8/2023

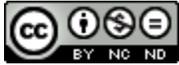

# Chapter 1: Understanding Prompt Engineering

## 1.1: Introduction

Prompt engineering is the process of creating useful commands and questions to engage students and also generate valuable responses from an artificial intelligence engine such as ChatGPT. It involves crafting practical questions, instructions, or suggestions to stimulate learning and valuable feedback. These prompts catalyze students to explore new knowledge, analyze complex concepts, and reflect on their understanding, enhancing their learning experience.

Prompt engineering emphasizes the iterative and reflective nature of prompts. Prompting isn't a one-and-done activity but a continuously evolving process where prompts are designed, deployed, assessed, and refined over time.

In health care education, a thorough understanding of prompt engineering helps teachers be better educators and students learn the material faster and better. The healthcare curriculum can be both dense and complex. Mastering the subjects demands both theoretical knowledge and practical application. Well-designed prompts can help by facilitating engagement with the topic, promoting critical thinking, and boosting retention.

## 1.2: The Importance of Prompt Engineering in Education

Prompt engineering helps improve engagement, understanding, retention, and critical thinking. Teachers utilizing good prompts for students maximize educational time. Students querying AI engines such as ChatGPT or Google Bard will get more accurate and useful results with sound prompt engineering.

**Understanding:** good prompts illuminate complex ideas, facilitate a deeper understanding, and enable students to connect theory with practice.

**Retention:** Prompts improve active recall. Students can better understand and retain the material by repeatedly posing questions from different angles. Sound prompt engineering helps move knowledge from short-term to long-term memory.

**Critical Thinking:** Good prompts can stimulate crucial thinking skills. Prompts help students see the subject from various viewpoints, helping them analyze, evaluate, and synthesize information. Similarly, good prompting of AI engines generates a wide range of responses which can help the user get better results from AI tools.

**Clinical Applications:** Prompts help medical and nursing students apply their theoretical knowledge to real-world situations. With good prompts, students have improved decision-making and problem-solving skills. Prompt engineering helps users by getting AI engines to respond with real-world clinical scenarios.

## 1.3: Basic Concepts and Terminology

**Prompt:** a prompt is a question, statement, or instruction that triggers a specific response when posed by a teacher to the student or when posed by a human user to an AI engine such as ChatGPT. Prompts inspire learning, discussion, reflection, or action in educational settings.

**Prompt Engineering:** creating and refining prompts to get a helpful response. Typically, prompt engineering refers to the refinement of prompts given by a human user to an AI. However, prompt engineering also plays a vital role in the education of medical and nursing students. Prompt engineering creates good prompts given by teachers to students to stimulate thinking and learning. Engineering involves designing, implementing, and refining prompts to maximize their effectiveness in eliciting desired responses or behaviors.

**Closed-Ended Prompts:** prompts that call for a definitive answer are closed-ended. These types of prompts expect a factual response to the specific question. Such prompts assess a student's understanding of basic concepts. When presented to an AI, the prompt will return a fact-based answer.

**Open-Ended Prompts:** these prompts encourage more expansive, thoughtful, and sometimes theoretical responses designed to stimulate higher-order thinking. They're helpful when exploring complex topics, promoting analysis, or fostering creativity.

**Socratic Prompts:** these prompts are a series of questions after the method used by the ancient Greek philosopher Socrates. They encourage critical thinking by asking questions that each build upon the previous question to probe deeper into the topic.

**Reflective Prompts:** when to goal is to stimulate introspection, self-awareness, and personal growth, reflective prompts can be utilized. They're often used in journaling or self-evaluation activities.

**Student / AI:** the student may be the human the teacher prompts or a user's AI engine. At the time of this writing, the most common AI engines are ChatGPT and Google Bard. Because the principles of prompt engineering in the classroom are identical to the principles of prompt engineering for AI engines, the terms "student" and "AI" will be used interchangeably throughout this book.

## 1.4 Questions

1. Define prompt engineering in your own words and explain its relevance to medical and nursing education.
2. Describe an instance from your medical or nursing education where a well-constructed prompt facilitated your understanding of a complex concept.
3. Reflect on the cognitive processes triggered when you respond to an instructional prompt.

4. Given the importance of cognitive science in prompt engineering, how can understanding cognitive processes improve the creation of effective prompts?
5. How can prompt engineering help foster critical thinking and self-directed learning in medical and nursing students?
6. Discuss the limitations of traditional teaching methods in medical and nursing education and how prompt engineering can address these limitations.
7. Consider the role of Socratic questioning in medical and nursing education. How can it be applied in modern classroom settings?
8. Draw parallels between the principles of active learning and the application of prompt engineering in medical and nursing education.
9. Imagine you are explaining the importance of prompt engineering to a peer who is unfamiliar with the concept. How would you convey its value in medical and nursing education?
10. Consider a topic or concept in your field of study that students often find challenging. How might you utilize prompt engineering to facilitate better understanding and engagement with this topic?

## 1.5 Key Takeaways

1. Prompt engineering is an iterative process of creating and then refining questions that elicit the desired response from students or an AI.
2. Given the complexity of medicine, well-engineered prompts can improve student engagement, understanding, memory, and critical thinking. Good prompts help prepare students for clinical practice.
3. The different types of prompts include closed-ended prompts (fact-based), open-ended prompts (encourage expansive thinking), Socratic prompts (stimulate critical thinking), and reflective prompts (foster introspection and personal growth).
4. Prompt engineering is the dynamic process of continuous and never-ending improvement, with the goal being effective prompts that obtain good feedback.

## 1.6: Example Prompts for Chapter 1

1. What are the three major types of prompts? What is each type used for? Give an example of each prompt type, then test and refine it.
2. How do prompts promote critical thinking?
3. What are the benefits of using reflective prompts? How should each subsequent prompt build upon the previous one? What is the result?
4. How can prompt engineering improve learning?
5. Design an effective prompt for medical or nursing students. Create a prompt that results in a helpful reply from an AI engine. Test it out on ChatGPT or Google Bard.

# Chapter 2: Anatomy of a Good Prompt

Good prompts are thought out in advance. They are not simply stream of thought questions. Good prompts are intentionally crafted after thought, contemplation, and testing. A good prompt will provoke curiosity, improve learning, and help to evaluate a topic thoroughly.

## 2.1: Clear and Precise Input

The primary qualities of a good prompt are clarity and precision. These are required to obtain the desired response, avoid confusion, and keep the learner engaged. A clear prompt has very little ambiguity.

Clarity is the clearness of the prompt. A prompt that has clarity is easy to understand. The effective prompt will have clear instructions.

Precision is a process that involves specificity. A precise prompt focuses on a specific aspect of a topic, focusing the question in a well-defined manner. Precise prompts keep the discussion focused, avoiding overly broad and generalized discussions.

For example, the prompt "Discuss diabetes" is too vague. The result is that students will need clarification. No starting point is given. Such a prompt needs to be more transparent and precise to improve learning. A better prompt would be "analyze the role of insulin resistance in developing Type 2 Diabetes" because the intent is clear and the focus is precise. This guides the AI in a specific avenue of exploration.

## 2.2: Relevance to Learning Objectives

The prompt should serve a clear purpose in the learning journey, whether understanding a concept, applying a theory, or developing a skill. Every prompt should align with one or more learning objectives. This ensures students' engagement with the prompt contributes to their progress and development.

To maintain relevance, prompts should be derived from the core curriculum and tailored to the level of the student's knowledge and skills. For example, a prompt on advanced cardiac life support would be relevant for final-year nursing students preparing for critical care roles but inappropriate for first-year students still learning human anatomy.

## 2.3: Stimulating Critical Thinking

While recall-based prompts have their place in knowledge reinforcement, good prompts often go beyond mere memory retrieval. They stimulate critical thinking, encouraging students to analyze, evaluate, or synthesize information.

In medical and nursing education, these critical thinking prompts involve interpreting patient symptoms, comparing treatment options, predicting disease progression, or discussing ethical issues in healthcare. Such prompts promote a deeper understanding of the subject matter and hone the higher-order cognitive skills essential for healthcare professionals.

## 2.4: Incorporating Practical Application

Medical and nursing education is not just about theoretical knowledge; practical application forms its backbone. Good prompts often bridge the gap between theory and practice, encouraging students to apply their knowledge to real-life or simulated scenarios.

These prompts could involve case studies, role-plays, or decision-making scenarios that reflect the challenges faced in clinical practice. For instance, a prompt could present a patient case and ask the students to devise a nursing care plan or discuss the medical management strategy. Such prompts provide invaluable opportunities for students to practice their clinical reasoning skills and prepare for their professional roles.

## 2.5: Provoking Reflection and Self-Assessment

Lastly, a good prompt often provokes reflection and self-assessment. Reflective prompts engage students in introspection about their learning process, performance, or experiences. They facilitate self-awareness and lifelong learning habits, key for medical and nursing professionals who must continuously adapt to evolving healthcare landscapes.

For instance, after a clinical rotation or a simulation session, prompts could ask students to reflect on what they did, why they did it, what went well, what didn't, and how to improve. Such reflective practice fosters personal growth and professional development.

In conclusion, a good prompt combines clarity, precision, relevance, critical stimulation, practical applicability, and reflective provocation. However, the exact blend of these elements can vary depending on the learning context, objectives, and the student's level. As we proceed further into this guide, we'll delve into the different types of prompts and how they can be engineered for various learning scenarios in medical and nursing education.

## 2.6: Questions

1. Describe a time when a well-crafted prompt led to a significant learning experience for you in medical or nursing education.
2. Reflect on the elements that contribute to making a prompt 'good.' How do these elements interact to facilitate learning?
3. Consider a prompt you've recently encountered that needed to be more effective or transparent. How would you re-design it based on the principles of good prompt anatomy?
4. How do the components of a good prompt align with the cognitive and pedagogical theories discussed in Chapter 1?
5. Think about a complex concept in your medical or nursing studies. Design a prompt that could facilitate a deep understanding of this concept. Explain your choice of prompt structure and design.

6. Discuss how the context (e.g., the learners, the learning environment, the subject matter) can influence the design of a good prompt.
7. In the context of nursing education, how would you design a reflective prompt to encourage students to contemplate the ethical implications of a clinical decision?
8. Analyze how a good prompt can foster active learning and critical thinking in medical or nursing students.
9. How can the Socratic method inform the design of effective prompts in medical and nursing education?
10. Imagine you're teaching a complex clinical procedure to medical students. Develop a set of prompts to guide the students' learning before, during, and after the procedure.

## 2.8: Key Takeaways

1. **Clarity and Precision:** Good prompts are clear and precise, free from ambiguity and vagueness. They use simple, direct language and target a specific aspect of a topic to focus the learner's attention effectively.
2. **Relevance to Learning Objectives:** An effective prompt aligns with the learning objectives. It serves a clear purpose in the learning journey, ensuring that engagement with the prompt contributes directly to the learner's progress.
3. **Stimulation of Critical Thinking:** Good prompts often go beyond memory retrieval and stimulate critical thinking. They encourage learners to analyze, evaluate, or synthesize information, promoting a more profound understanding and honing higher-order cognitive skills.
4. **Incorporation of Practical Application:** Effective prompts bridge the gap between theory and practice, enabling learners to apply their knowledge to real-world or simulated scenarios. Such prompts foster clinical reasoning skills and readiness for professional roles.
5. **Provocation of Reflection and Self-Assessment:** Reflective prompts provoke introspection and self-assessment, fostering self-awareness, lifelong learning habits, and professional development. They are especially vital in medical and nursing education, where continuous adaptation to evolving landscapes is essential.

*Tips for writing effective prompts:*

1. Use simple, direct language.
2. Avoid overly complex vocabulary.
3. Structure the prompt in a manner that makes the expected response clear.
4. Consider the learning objectives when writing the prompt.
5. Use a variety of prompt types to stay motivated to learn

## 2.9: Example Prompts for Chapter 2

1. **Clarity and Precision:**
    a. What are the primary physiologic differences between Type 1 and Type 2 diabetes? Use a scientific tone in your response.
    b. Describe how insulin resistance can lead to Type 2 diabetes in an overweight 50-year-old. What does insulin resistance mean? How is insulin resistance treated?
2. **Relevance to Learning Objectives:**
    a. Discuss the importance of handwashing in preventing the spread of infection.
    b. Analyze the ethical implications of physician-assisted suicide.
3. **Stimulating Critical Thinking:**
    a. Compare and contrast the different treatment options for hypertension.
    b. Predict how chronic obstructive pulmonary disease will likely progress in a middle-aged current smoker with a pack history of 30 years.
4. **Incorporating Practical Application:**
    a. Devise a nursing care plan for a patient with pneumonia.
    b. Discuss the medical management strategy for a patient with heart failure.
5. **Provoking Reflection and Self-Assessment:**

a. Reflect on your experiences during your most recent clinical rotation. What did you learn? What lasting lessons did you get out of it?
b. What did you do well on your last clinical rotation? What can you improve? What are your goals?

# Chapter 3: Different Types of Prompts

Understanding the different types of prompts is crucial for developing a diverse and flexible approach to learning in medical and nursing education. Other kinds of prompts stimulate different forms of thinking, ensuring a well-rounded and comprehensive learning experience. This chapter explores various prompts, each with its unique strengths and applications.

## 3.1: Closed-Ended Prompts

Closed-ended prompts are designed to elicit a specific response, often based on recall or comprehension. They usually have a definitive answer, making them ideal for assessing factual knowledge or basic understanding.

For instance, a closed-ended prompt in anatomy might ask, "What are the four chambers of the human heart?" or "Identify the bones in the human forearm."

While these prompts might not stimulate higher-order thinking, they reinforce foundational knowledge, provide quick feedback, and boost memory retention through active recall.

## 3.2: Open-Ended Prompts

Unlike closed-ended prompts, open-ended prompts encourage expansive, thoughtful responses. These prompts typically start with 'how', 'why', 'what', 'discuss', or 'explain', and stimulate higher-order thinking skills such as analysis, synthesis, evaluation, and creativity.

In a medical context, an open-ended prompt might ask, "Discuss the physiological changes during pregnancy" or "Explain how chronic hypertension affects the cardiovascular system."

Open-ended prompts are perfect for deepening understanding, promoting critical thinking, fostering creativity, and facilitating meaningful discussions. However, they might require more time to answer and assess.

## 3.3: Role-Play Prompts

Role-play prompts involve students acting out roles or scenarios, often based on real-world or hypothetical situations. These prompts simulate thinking of an issue from different perspectives. They can help the student build empathy, improve communication, and make better decisions.

In nursing education, role-play prompts could include scenarios such as, "You're a nurse in an emergency department, and a patient comes in complaining of severe chest pain. What do you do?" or "Role-play a conversation with a patient who is non-compliant with their medication."

Role-play prompts are particularly effective in preparing students for healthcare professions' interpersonal and practical challenges. However, they require a safe, supportive environment to be conducted effectively.

## 3.4: Socratic Prompts

Named after the ancient Greek philosopher Socrates, Socratic prompts involve questions designed to stimulate critical thinking and illuminate truth. These prompts encourage students to question their assumptions, examine their beliefs, and reason out their answers.

Socratic questioning might look like this in a clinical context: "What are the possible causes of this patient's symptoms? What evidence supports each possibility? What else do we need to know to make a definitive diagnosis?"

Socratic prompts promote thorough exploration of topics, improve reasoning skills, and foster a culture of curiosity and inquiry. However, they require skillful facilitation to ensure a productive and respectful dialog.

## 3.5: Reflective Prompts

Reflective prompts engage students in introspection about their learning process, experiences, or performance. These prompts facilitate self-awareness, self-assessment, and personal growth.

For instance, after a clinical rotation, a reflective prompt might ask, "Reflect on your interaction with patients today. What went well, and what challenges did you face? How will these experiences influence your future practice?"

Reflective prompts are integral to developing lifelong learning habits, emotional intelligence, and professional competencies. They are commonly used in portfolios, journals, or self-evaluation activities.

In conclusion, different types of prompts cater to various learning objectives and stimulate divergent forms of thinking. A good educator (or self-learner) is a prompt engineer who uses multiple prompts to facilitate an enriching learning experience.

## 3.6: Questions

1. List and briefly explain three types of prompts that you have encountered in your medical or nursing education.
2. Reflect on the advantages and disadvantages of recall prompts. When should they be used effectively in medical and nursing education?
3. Design a concept mapping prompt to help nursing students understand the connection between patient lifestyle factors and the development of chronic diseases.
4. Discuss how reflective prompts can contribute to developing empathy and professionalism in medical students.
5. Imagine you're teaching a class on diagnosing common pediatric conditions. Create a series of Socratic prompts to guide students in their decision-making process.
6. Choose a complex medical or nursing concept. Develop three prompts (e.g., recall, conceptual, reflective) to facilitate learning about this concept.
7. Reflect on a significant clinical experience you had. How could a reflective prompt help you process and learn from this experience?
8. Describe a situation where a procedural prompt could be beneficial in teaching a clinical skill.
9. Analyze how different types of prompts stimulate other cognitive processes.
10. Consider a medical or nursing education topic that often generates debate or differing opinions. How might a discussion prompt facilitate a productive and respectful debate among students?

## 3.7: Key Takeaways

1. **Understanding Various Prompt Types:** Different types of prompts stimulate various forms of thinking, fostering a comprehensive learning experience. They can be classified into closed-ended, open-ended, role-play, Socratic, and reflective prompts.
2. **Closed-Ended Prompts:** These prompts are designed to elicit specific responses based on recall or comprehension. They are essential for reinforcing foundational knowledge and providing immediate feedback.

3. **Open-Ended Prompts:** Open-ended prompts stimulate higher-order thinking skills such as analysis, synthesis, evaluation, and creativity. They deepen understanding and encourage critical thinking.
4. **Role-Play and Socratic Prompts:** Role-play prompts provide interactive, hands-on learning experiences fostering practical skills like communication and decision-making. Socratic prompts, on the other hand, stimulate critical thinking, reasoning, and curiosity.
5. **Reflective Prompts:** These prompts engage students in introspection about their learning process or experiences, facilitating self-awareness, self-assessment, and personal growth. They are crucial for fostering lifelong learning habits and professional competencies.

## 3.9: Example Prompts for Chapter 3

**Closed-Ended Prompts**

- What are the four chambers of the human heart?
- Identify the bones in the human forearm.
- List the three primary types of human blood cells.
- What are the five stages of cellular respiration?
- What are the three main types of muscle tissue?

**Open-Ended Prompts**

- Discuss the physiological changes during pregnancy.
- Explain how chronic hypertension affects the cardiovascular system.
- What are the ethics and morality of physician-assisted suicide?
- How can we improve a hospitalized patient's experience and satisfaction?
- What are the future trends in healthcare?

**Role-Play Prompts**

- You are an RN working in a busy emergency department. A patient comes in complaining of severe substernal chest pain. What do you do first? Then what?
- Role-play a conversation with a patient who is non-compliant with their medication.
- You are a doctor and have to break bad news to a patient. How do you do it?
- You are the hospital's patient advocate. You become aware of a patient that appears to be getting substandard treatment. What do you do first? Then what?
- You are a healthcare administrator and must decide how to allocate resources. What do you do?

**Socratic Prompts**

- A 60-year-old has stomach pain that gets worse with exercise. Give a differential diagnosis with evidence to support each one. What else do we need to know? Which diagnosis requires immediate action? What is the worst diagnosis that must be ruled out immediately?
- What are the ethical implications of this research study? What are the potential benefits and risks to the patient? How can we minimize the risks and increase the benefits?
- How can we view [diagnosis] from different perspectives? What biases are at play? List 5 different biases that health care professionals can have when trying to come up with an accurate diagnosis.
- Are there multiple ways to treat [condition]? What is the standard of care? What are the advantages and disadvantages of other approaches? How can we effectively work with patients that choose an alternative medicine approach?
- How can we improve our process of taking a history and physical? List 5 ways to improve the process. What are the potential benefits and risks of each proposal?

**Reflective Prompts**

- Reflect on your interaction with patients today. What went well, and what challenges did you face? How will these experiences influence your future practice?

- Reflect on your learning process for this course. What were the most challenging concepts? What were the most rewarding experiences? How can you implement the lessons that you have learned?
- Reflect on your personal growth and development over the past year. What did you do well? What can you improve? How has this past year changed your goals? What challenges ahead do you foresee?

# Chapter 4: Application

Having established an understanding of the diverse types of prompts and their characteristics, we now apply these prompts effectively in the context of medical and nursing education. Each section will focus on a distinct aspect of these fields, illustrating how prompt engineering can facilitate meaningful and compelling learning experiences.

## 4.1: Anatomy and Physiology

The foundational pillars of medical and nursing knowledge, anatomy, and physiology, require a deep understanding of the human body's structure and functions. Closed-ended prompts reinforce the basic facts of these subjects, e.g., "Identify the major organs of the digestive system" or "List the types of blood cells and their functions."

Open-ended prompts can help deepen understanding and foster connections, such as, "Explain the impact of aging on the cardiovascular system" or "Discuss the interplay between the nervous and endocrine systems in managing stress."

## 4.2: Pathology and Pharmacology

Understanding disease processes and drug actions are vital areas where prompts can be effectively utilized. Closed-ended prompts can assess basic comprehension, e.g., "What are the stages of cancer development?" or "What is the mechanism of action of beta-blockers?"

Open-ended and Socratic prompts can promote critical thinking and integration of knowledge. For example, "How does an autoimmune disorder develop, and how can it be managed?" or "Discuss the effects and side effects of opioid analgesics. How would you manage a patient on these medications?"

## 4.3: Clinical Skills

Developing clinical skills requires practice and application. Role-play prompts can simulate clinical scenarios and foster decision-making, communication, and technical skills. For example, "You are the first responder to a patient who collapsed on the ward. Act out your response," or "Role-play a conversation with a patient newly diagnosed with diabetes."

Socratic prompts can facilitate reflection and learning during and after skill demonstrations, e.g., "Why did you choose this approach for wound dressing? What other methods could you have used?"

## 4.4: Patient Assessment and Management

Patient assessment and management form the crux of medical and nursing practice. Open-ended, Socratic, and role-play prompts can be used to develop and evaluate these skills.

For example, "Given this patient's history, symptoms, and vital signs, what are the potential diagnoses, and what further investigations would you order?" or "Role-play breaking bad news to a patient," or "What factors would you consider in creating a care plan for a postoperative patient?"

## 4.5: Ethics, Communication, and Professionalism

In these soft but crucial areas, reflective and role-play prompts are instrumental. For example, "Reflect on an ethical dilemma you encountered or read about. How would you handle it?" or "Role-play dealing with a difficult family member or colleague."

Reflective prompts can also facilitate continuous professional development, e.g., "Reflect on your performance during the recent clinical rotation. What strengths did you exhibit, and what areas do you need to improve?"

## 4.6: Lifelong Learning

Medical and nursing professions require continuous learning and adaptation. Reflective prompts can foster self-directed learning, e.g., "Reflect on your learning goals for the next year" or "Review a recent research article in your field and reflect on its implications for your practice."

In conclusion, prompt engineering is a versatile medical and nursing education tool enhancing engagement, understanding, skill development, reflection, and lifelong learning. The application of prompts is limited only by the educator's creativity and the learner's curiosity, allowing for rich and diverse learning experiences tailored to the demands and challenges of the healthcare field.

## 4.7: Questions

1. Recall a lecture from your medical or nursing education where the instructor effectively integrated prompts. What impact did these prompts have on your learning?
2. Discuss how integrating prompts into lectures can enhance the learning experience in medical or nursing education.
3. Consider a topic that you find challenging in your studies. How could integrated prompts within a case study help you better understand this topic?
4. Reflect on your experiences with simulation-based learning. How can prompts be effectively used in this context to improve skill acquisition and decision-making?
5. Design a series of prompts that could be integrated into a group project on public health promotion.
6. Discuss how integrating prompts into e-learning platforms can enhance self-directed medical or nursing education learning.
7. Reflect on the role of prompts in clinical rotations or practical placements. How do they contribute to the transition from theory to practice?

8. Consider a complex surgical procedure. How could you integrate procedural prompts into a simulation training session for this procedure?
9. Evaluate how different types of prompts (e.g., recall, reflective, procedural) can be integrated into various learning activities (e.g., lectures, case studies, simulations, group projects) to facilitate diverse learning outcomes.
10. How can prompts foster effective communication and collaboration between medical and nursing students in the context of interprofessional education?

## 4.8: Key Takeaways

1. **Anatomy and Physiology:** Closed-ended prompts reinforce basic facts, while open-ended prompts deepen understanding and foster connections in the foundational subjects of anatomy and physiology.
2. **Pathology and Pharmacology:** Closed-ended prompts assess basic comprehension, while open-ended and Socratic prompts promote critical thinking and integration of knowledge in understanding disease processes and drug actions.
3. **Clinical Skills:** Role-play prompts simulate clinical scenarios to foster decision-making, communication, and technical skills. Socratic prompts facilitate reflection and learning during and after skill demonstrations.
4. **Patient Assessment and Management:** Open-ended, Socratic, and role-play prompts are used to develop and evaluate patient assessment and management skills, which form the crux of medical and nursing practice.
5. **Ethics, Communication, and Professionalism:** Reflective and role-play prompts are instrumental in fostering ethics, communication skills, and professionalism, which are crucial "soft" skills in medicine and nursing.
6. **Lifelong Learning:** Reflective prompts foster self-directed learning, which is necessary for continuous learning and adaptation, a cornerstone of the medical and nursing professions.

# 4.9: Example Prompts for Chapter 4

1. **Anatomy and Physiology**
   a. Closed-ended: Identify the major organs of the digestive system.
   b. Open-ended: Explain the impact of aging on the cardiovascular system.
   c. Socratic: How does the nervous system coordinate the body's movements?
2. **Pathology and Pharmacology**
   a. Closed-ended: What are the stages of cancer development?
   b. Open-ended: How does an autoimmune disorder develop, and how can it be managed?
   c. Socratic: What are the potential side effects of opioids for pain relief?
3. **Clinical Skills**
   a. Role-play: You are the first responder to a patient who collapsed in the ward. Act out your response.
   b. Socratic: Why did you choose this approach for wound dressing? What other methods could you have used?
   c. Reflective: What did you learn from practicing this skill?
4. **Patient Assessment and Management**
   a. Open-ended: Given this patient's history, symptoms, and vital signs, what are the potential diagnoses, and what further investigations would you order?
   b. Role-play: Break devastating news to a patient.
   c. Reflective: What factors would you consider in creating a care plan for a postoperative patient?
5. **Ethics, Communication, and Professionalism**
   a. Reflective: Reflect on an ethical dilemma you encountered or read about. How would you handle it?
   b. Role-play: Deal with a difficult family member or colleague.
   c. Reflective: What are your strengths and weaknesses as a communicator?

# Chapter 5: Strategies for Effective Prompt Design and Utilization

After understanding the different types of prompts and their applications in medical and nursing education, the next step is to master the art of prompt design and utilization. This chapter will offer practical strategies for creating and deploying effective prompts to facilitate learning and skill development.

## 5.1: Define Your Learning Objectives

Before designing a prompt, it is crucial to establish your learning objectives. What knowledge or skill do you want the students to gain or demonstrate? What level of understanding or proficiency are you targeting? The answers to these questions will guide your prompt type and content choice. For example, a closed-ended prompt might be appropriate if your objective is to assess students' basic knowledge of human anatomy. However, an open-ended or Socratic prompt might be more fitting if you aim to stimulate critical thinking about disease mechanisms, an open-ended or Socratic prompt might be more relevant.

## 5.2: Consider the Learners' Level

Design prompts appropriate for your student's knowledge, skill, and experience level. A prompt that is too easy might not stimulate learning, while an excessively difficult prompt can lead to frustration or disengagement. A good prompt should challenge but not overwhelm the learners. Moreover, as students progress in their education, the prompts should evolve, becoming more complex and integrating more aspects of their learning.

## 5.3: Design for Clarity and Precision

As discussed in Chapter 2, clarity and precision are critical attributes of an effective prompt. Use simple, direct language, structure your prompt clearly, and specify what you want the students to do. Avoid ambiguous or vague terms that could confuse the students. A well-crafted prompt should leave no room for misunderstanding about what is expected of the students.

## 5.4: Create a Safe and Supportive Learning Environment

The learning environment often influences the effectiveness of prompts. For prompts to facilitate meaningful discussions, reflections, or role-plays, students must feel safe, respected, and supported. Foster an atmosphere of mutual respect, openness, and constructive feedback. Encourage students to voice their thoughts, questions, or concerns and validate their contributions.

## 5.5: Use Prompt Sequencing and Layering

Prompt sequencing involves using a series of prompts in a logical and progressive order to guide learning. For example, you might start with closed-ended prompts to establish fundamental knowledge, then proceed to open-ended prompts to stimulate analysis or synthesis, and finally use reflective prompts to facilitate self-assessment or future planning.

Prompt layering involves using multiple prompts within a single learning activity or session. For instance, in a case study discussion, you could use an open-ended prompt to start the discussion, Socratic prompts to probe deeper into the case, a role-play prompt to simulate a clinical decision, and a reflective prompt to wrap up the session.

## 5.6: Provide Constructive Feedback

Feedback plays a crucial role in the learning process. Whether for a closed-ended prompt where you correct an incorrect response or for an open-ended or reflective prompt where you provide comments or suggestions, make your feedback constructive, specific, and timely. Use feedback to correct errors and affirm correct answers, acknowledge effort, and encourage deeper thinking or self-improvement.

## 5.7: Review and Revise Your Prompts

Finally, prompt engineering is a continuous learning, experimentation, and refinement process. Regularly review and revise your prompts based on your observations, students' responses, and feedback from students or colleagues. Are the prompts achieving their intended learning objectives? Are they engaging the students effectively? Are there any recurring issues or challenges? The answers to these questions can guide your prompt refinement and development process.

In conclusion, prompt engineering is both a science and an art, involving a thoughtful blend of pedagogical principles, practical considerations, and creative innovation. Mastering this skill can significantly enhance the effectiveness of your teaching or self-learning in medical and nursing education. As you embark on this exciting journey, remember to keep your learners at the heart of your prompt design and utilization strategies. After all, the ultimate goal of prompt engineering is to facilitate meaningful, learner-centered education that prepares future healthcare professionals for the challenges and rewards of their noble profession.

## 5.8: Questions

1. Discuss how technology can enhance the effectiveness of prompts in medical and nursing education.
2. Reflect on an instance where a technology-enhanced prompt significantly improved your learning experience.
3. Consider a topic that medical or nursing students often struggle with. How could an AI-powered adaptive learning system use prompts to support learning in this area?
4. How might virtual reality (VR) technology be used to deliver effective prompts in the context of simulation-based learning?
5. Discuss the potential benefits and challenges of using AI to personalize prompts based on individual learners' needs and progress.

6. Evaluate how data analytics can inform the design and delivery of prompts in digital learning environments.
7. How can technology-enhanced prompts foster active learning and critical thinking in medical and nursing education?
8. Design a series of prompts for a mobile learning app aimed at teaching pharmacology to nursing students.
9. Discuss the ethical considerations when using AI and data analytics in prompt engineering.
10. Reflect on the future of prompt engineering in medical and nursing education. How do you envision technology and AI transforming the way prompts are used?

## 5.9: Key Takeaways

1. **Define Your Learning Objectives:** Before designing a prompt, establish your learning objectives. What you want the students to gain or demonstrate will guide your prompt type and content choice.
2. **Consider the Learners' Level:** Design prompts appropriate for your student's knowledge, skill, and experience level. A good prompt should challenge but not overwhelm learners.
3. **Design for Clarity and Precision:** Use simple, direct language, and structure your prompt. An effective prompt should leave no room for misunderstanding.
4. **Create a Safe and Supportive Learning Environment:** For prompts to facilitate meaningful learning, students need to feel safe, respected, and supported. Encourage openness, respect, and constructive feedback in the learning environment.
5. **Use Prompt Sequencing and Layering:** Prompt sequencing involves using prompts in a logical and progressive order, while prompt layering uses multiple prompts within a single learning activity.
6. **Provide Constructive Feedback:** Feedback plays a crucial role in learning. Make your feedback constructive, specific, and timely, affirming correct answers, acknowledging effort, and encouraging more profound thinking.

7. **Review and Revise Your Prompts:** The art of prompt engineering involves continuous learning, experimentation, and refinement. Review and revise your prompts based on observations, student responses, and feedback.

## 5.10: Example Prompts for Chapter 5

1. **Closed-ended prompt:** What are the four chambers of the human heart?
2. **Open-ended prompt:** Discuss the physiological changes during pregnancy.
3. **Socratic prompt:** What are the possible causes of this patient's symptoms? What evidence supports each possibility? What else do we need to know to make a definitive diagnosis?
4. **Role-play prompt:** You are a nurse in an emergency department, and a patient complains of severe chest pain. What do you do?
5. **Reflective prompt:** Reflect on your learning process for this course. What were the most challenging concepts? What were the most rewarding experiences? How can you apply what you have learned in this course to your future practice?

# Chapter 6: Adapting Prompt Engineering for Digital Learning Environments

As medical and nursing education increasingly adopts digital technologies, there is a growing need to adapt our teaching and learning strategies for these new environments. This chapter will focus on how we can adapt and optimize prompt engineering techniques for digital learning platforms, ensuring we continue offering an engaging and practical learning experience.

## 6.1: The Digital Learning Environment

Digital learning environments are online or digital platforms facilitating education, such as virtual classrooms, learning management systems, discussion forums, mobile apps, and virtual reality platforms. Digital learning offers unique advantages, including flexibility, scalability, accessibility, and the potential for personalized learning. However, it presents unique challenges like technical issues, learner isolation, and engagement difficulties. Prompt engineering can play a crucial role in maximizing the benefits and addressing the challenges of digital learning.

## 6.2: Adapting Prompts for Digital Text-Based Platforms

Digital text-based platforms, such as online forums, chats, or email threads, are common in digital learning. These platforms are ideal for open-ended, reflective, and Socratic prompts that stimulate discussion and reflection. They allow students to articulate their thoughts in writing, facilitating critical thinking and writing skills.

Prompts can be designed to initiate discussions, deepen the conversation, and guide the learners toward meaningful insights. For instance, an initial prompt might ask students to discuss a case study or article. In contrast, follow-up prompts could probe deeper into the issues, elicit diverse perspectives, or encourage students to connect the topic with their experiences or future practice.

## 6.3: Adapting Prompts for Synchronous Online Sessions

Synchronous online sessions, such as webinars or virtual classes, offer real-time interaction like traditional classrooms. These sessions can utilize a wide range of prompts, including closed-ended prompts for instant polls or quizzes, open-ended prompts for group discussions or breakout sessions, role-play prompts for simulated scenarios, and reflective prompts for debriefings or wrap-ups.

Using multimedia (e.g., slides, videos, virtual whiteboards) and interactive features (e.g., chat, poll, raise hand) can enhance the engagement and effectiveness of prompts. Furthermore, digital tools can facilitate prompt responses and feedback, such as online quizzes for instant grading or chat/comments for written responses and feedback.

## 6.4: Adapting Prompts for Asynchronous Online Sessions

Asynchronous online sessions, such as pre-recorded lectures or self-paced courses, allow learners to engage with the content at their own pace. However, the lack of real-time interaction can make engagement and feedback more challenging.

Strategically placed prompts can help maintain learners' attention, deepen their understanding, and facilitate self-assessment. For instance, closed-ended prompts can be used for self-check quizzes or interactive elements within the lectures. Open-ended or reflective prompts can be used for post-lecture assignments or discussion forum posts.

## 6.5: Adapting Prompts for Mobile Learning

Mobile learning, through apps or mobile-optimized platforms, offers convenience and portability. Bite-sized learning materials, often in the form of flashcards or quizzes, can utilize closed-ended prompts for quick knowledge checks or reinforcement.

Moreover, mobile notifications can deliver timely prompts, such as daily reflection questions or challenge problems. Some apps also offer interactive case studies or scenarios, which can incorporate open-ended or role-play prompts for decision-making practice.

## 6.6: Adapting Prompts for Virtual Reality or Simulation Platforms

Virtual reality (VR) or simulation platforms offer immersive, experiential learning experiences. These platforms are ideal for role-play prompts that simulate clinical scenarios or procedures. They can also incorporate open-ended, Socratic, or reflective prompts for decision-making, problem-solving, debriefing, or reflection.

For example, a VR clinical scenario could prompt the learner to assess a virtual patient, make diagnostic and therapeutic decisions, and communicate with the patient or team. The scenario could also include prompts for reflection or debriefing about the performance, decision-making process, or learning points.

In conclusion, prompt engineering can be effectively adapted for digital learning environments, capitalizing on their unique features and advantages to facilitate engaging, interactive, and meaningful learning experiences. As digital learning continues to evolve, so too should our prompt engineering strategies, staying attuned to emerging technologies, learner needs, and pedagogical insights.

## 6.7: Questions

1. Reflect on a time when feedback from a prompt significantly influenced your learning process in medical or nursing education.

2. Discuss how immediate feedback from a prompt can enhance the learning experience in a simulation-based training scenario.
3. How can reflective prompts encourage self-assessment and lifelong learning in medical and nursing professionals?
4. Design a series of prompts for a case study that integrates feedback at various stages of the learning process.
5. Consider the role of peers in providing feedback on responses to prompts. How can this process enhance collaborative learning and critical thinking skills?
6. Evaluate the importance of educator reflection in the process of prompt engineering. How can reflection inform the iterative refinement of prompts?
7. Discuss how feedback from prompts can guide learners' self-directed study in a flipped classroom model.
8. Describe a clinical scenario where feedback from a procedural prompt could improve patient safety and care outcomes.
9. Reflect on a significant clinical experience. How could a reflective prompt and subsequent feedback help you learn from this experience?
10. Consider a complex topic in medical or nursing education. How could an AI-powered adaptive learning system use feedback and reflection prompts to support learning in this area?

## 6.8: Key Takeaways

1. **The Digital Learning Environment:** Digital learning environments like online platforms and mobile apps offer unique advantages and challenges. Prompt engineering can maximize the benefits and address the challenges of digital learning.
2. **Adapting Prompts for Digital Text-Based Platforms:** Digital text-based platforms are suitable for open-ended, reflective, and Socratic prompts. Prompts can initiate discussions, deepen conversations, and guide learners toward meaningful insights.

3. **Adapting Prompts for Synchronous Online Sessions:** Synchronous online sessions can utilize many prompts. Using multimedia and interactive features can enhance the engagement and effectiveness of prompts.
4. **Adapting Prompts for Asynchronous Online Sessions:** In asynchronous online sessions, strategically placed prompts can maintain learners' attention, deepen their understanding, and facilitate self-assessment.
5. **Adapting Prompts for Mobile Learning:** Mobile learning can use closed-ended prompts for quick knowledge checks or reinforcement. Mobile notifications can deliver timely prompts, and interactive case studies can incorporate open-ended or role-play prompts.
6. **Adapting Prompts for Virtual Reality or Simulation Platforms:** VR or simulation platforms are ideal for role-play prompts that simulate clinical scenarios or procedures. They can also incorporate open-ended, Socratic, or reflective prompts for decision-making, problem-solving, debriefing, or reflection.

## 6.9: Example Prompts for Chapter 6

1. Digital Text-Based Platforms
    a. Open-ended prompt: Discuss the ethical implications of physician-assisted suicide.
    b. Socratic prompt: What are the different perspectives [condition]? How can we reach a consensus? Is a consensus critical?
2. Synchronous Online Sessions
    a. Closed-ended prompt: What are the four chambers of the human heart?
    b. Open-ended prompt: Discuss the physiological changes during pregnancy.
    c. Role-play prompt: You are a nurse in an emergency department, and a patient complains of severe chest pain. What do you do first? What next?
3. Asynchronous Online Sessions
    a. Closed-ended prompt: What are the three main types of blood cells?

b. Open-ended prompt: How can we improve the patient experience in the healthcare setting?
   c. Reflective prompt: Reflect on your learning process for this course. What were the most challenging concepts? What were the most rewarding experiences?
4. Mobile Learning
   a. Closed-ended prompt: What are the five stages of cellular respiration?
   b. Open-ended prompt: What are the future trends in healthcare?
   c. Reflective prompt: What are your strengths and weaknesses as a communicator? How can you improve?
5. Virtual Reality or Simulation Platforms
   a. Role-play prompt: You are a doctor and have to break bad news to a patient. How do you approach this conversation?
   b. Open-ended prompt: What are the ethical implications of using virtual reality in healthcare?
   c. Reflective prompt: Reflect on your experience using virtual reality in healthcare. How can virtual reality be used to improve healthcare?

# Chapter 7: Evaluating the Effectiveness of Prompts in Medical and Nursing Education

Now that we've explored prompt engineering techniques and how they apply to various contexts, it's time to discuss a crucial element of the process: evaluation. It's imperative to assess the effectiveness of your prompts to ensure they're meeting educational objectives. This chapter will delve into methods for evaluating prompts and using feedback to enhance future prompts.

## 7.1: Why Evaluate Prompts?

Assessing the effectiveness of prompts serves multiple purposes: it gauges whether the prompt met its learning objective, provides information to improve future prompts, and enhances the learning process by identifying gaps in understanding or communication. Given the critical role of prompts in medical and nursing education, regular evaluation is essential for ongoing pedagogical success.

## 7.2: Evaluating Closed-Ended Prompts

Closed-ended prompts are typically easy to evaluate, thanks to their definitive answers. However, more than a simple right or wrong assessment is needed. Pay attention to patterns of wrong answers; they can indicate areas of confusion or misunderstanding. For instance, if many students miss a particular prompt, it might be that the material was unclear or the prompt itself was confusing. Use this information to guide future teaching and prompt design.

## 7.3: Evaluating Open-Ended, Reflective, and Socratic Prompts

Assessing open-ended, reflective, and Socratic prompts is more complex due to the diverse and subjective nature of the responses. Here, rubrics come in handy. A rubric is a scoring guide used to evaluate the quality of learners' constructed responses.

When designing a rubric, consider the following criteria: comprehension of material, clarity of expression, depth of analysis or reflection, relevance of response to the prompt, and creativity or originality. Each criterion should have a scale, typically from 1 (poor) to 5 (excellent). The rubric helps ensure a fair and objective assessment, even for subjective responses.

## 7.4: Evaluating Role-Play Prompts

Role-play prompts, often used to simulate clinical scenarios, require assessing knowledge and skills. Again, a rubric can be beneficial. Criteria may include medical knowledge, clinical reasoning, technical skills, communication skills, professionalism, and teamwork. Another consideration for role-play is the realism of the performance, reflecting the students' ability to apply knowledge and skills in a realistic context.

## 7.5: Using Feedback for Evaluation

Feedback from students can offer invaluable insights into the effectiveness of prompts. Ask students how they found the prompt: Was it clear? Did it stimulate thinking or learning? Was it too easy or difficult? What would they suggest for future prompts?

Feedback can be collected through anonymous surveys or open discussions. Create a safe and supportive environment where students feel comfortable sharing their views.

## 7.6: Continuous Improvement

Evaluation isn't a one-time process but an ongoing cycle of improvement. Use the insights from each round of evaluation to refine your prompts and teaching methods. Keep abreast of the latest medical and nursing education research and innovations and prompt engineering. Experiment with new prompt types or formats, and continually challenge your students and yourself.

In conclusion, evaluating the effectiveness of prompts is as critical as the design and application of prompts. It completes the learning cycle, providing essential insights for continual growth and improvement. Investing in this process ensures that your prompt engineering skills continue to evolve, serving the ultimate goal of excellence in medical and nursing education.

## 7.7: Questions

1. Discuss an instance where collaboration in prompt engineering enhanced your learning experience in medical or nursing education.
2. Reflect on how collaborative prompts can foster teamwork and communication skills in medical and nursing students.
3. How can collaboration in prompt engineering facilitate interprofessional education between medical and nursing students?
4. Design a collaborative prompt for a group project focused on developing a patient care plan.
5. Discuss the benefits and challenges of incorporating collaboration into prompt engineering.
6. Reflect on a time when peer feedback from a collaborative prompt led to a significant learning outcome.
7. How can collaborative prompts be utilized in an online or blended learning environment?

8. Evaluate how collaboration in prompt engineering can foster a sense of community and shared learning among medical and nursing students.
9. Consider a complex clinical case. How could a collaborative prompt facilitate the exchange of ideas and perspectives to reach an optimal solution?
10. Reflect on the role of the educator in guiding collaborative prompts. How can they ensure effective collaboration and equitable participation among learners?

## 7.8: Key Takeaways

1. **Why Evaluate Prompts?:** Evaluating prompts helps gauge whether the prompt met its learning objective, improves future prompts, and enhances learning by identifying gaps in understanding or communication.
2. **Evaluating Closed-Ended Prompts:** Pay attention to patterns of wrong answers to closed-ended prompts, as they can indicate areas of confusion or misunderstanding. Use this information to guide future teaching and prompt design.
3. **Evaluating Open-Ended, Reflective, and Socratic Prompts:** The diverse and subjective nature of responses to these prompts requires a rubric for evaluation. Criteria can include comprehension of material, clarity of expression, depth of analysis or reflection, relevance of response to the prompt, and creativity or originality.
4. **Evaluating Role-Play Prompts:** Role-play prompts require assessing both knowledge and skills. Rubrics can be used for this evaluation, with criteria that may include medical knowledge, clinical reasoning, technical skills, communication skills, professionalism, and teamwork.
5. **Using Feedback for Evaluation:** Feedback from students provides invaluable insights into the effectiveness of prompts. Questions can be asked about clarity, whether the prompt stimulated thinking or learning and its difficulty level.
6. **Continuous Improvement:** Evaluation should be an ongoing cycle of improvement. Use insights from each round of evaluation to refine prompts and teaching methods.

Keep abreast of the latest research and innovations in the field, and experiment with new prompt types or formats.

## 7.9: Example Prompts for Chapter 7

1. Evaluating Closed-Ended Prompts
    a. What are the four chambers of the human heart?
    b. What are the three main types of blood cells?
    c. What are the five stages of cellular respiration?
    d. What are the future trends in healthcare?
2. Evaluating Open-Ended, Reflective, and Socratic Prompts
    a. Discuss the ethical implications of physician-assisted suicide.
    b. Reflect on your learning process for this course. What were the most challenging concepts? What were the most rewarding experiences?
    c. Does virtual reality have an essential role in healthcare? What are the potential benefits to patients? Could patients be harmed?
3. Evaluating Role-Play Prompts
    a. You are a doctor and must let the family know about the patient's passing. How do you approach this?
    b. You are a nurse in a busy family medicine clinic. Your next patient comes in complaining of abdominal pain. What is your most important next step?
    c. You are a patient advocate. A nurse tells you about a patient being mistreated. What do you do first? What is your thought process?
    d. You are a healthcare administrator and must decide how to allocate resources.
4. Using Feedback for Evaluation
    a. How did you find the prompts in this course? Were they clear? Did they stimulate thinking or learning? Were they too easy or difficult? What would you suggest for future prompts?
    b. Please rate the following prompts on a scale of 1 to 5 (1 = poor, 5 = excellent):

i. Clarity
                    ii. Stimulating
                    iii. Challenging
                    iv. Relevant
                    v. Thought-provoking
    5. Continuous Improvement
        a. What are your plans for continuous improvement in your prompt engineering skills?
        b. What new research or innovations in medical and nursing education and prompt engineering are you interested in learning more about?
        c. What new prompt types or formats would you like to experiment with?
        d. How can you continually challenge yourself and your students to improve their learning?

# Chapter 8: Incorporating Advanced AI Technology into Prompt Engineering

As we embrace the digital age, artificial intelligence (AI) is increasingly becoming an integral part of our lives, including education. AI-powered applications can augment teaching and learning in medical and nursing education, particularly in prompt engineering. This chapter will explore using advanced AI technologies for designing, deploying, and evaluating prompts.

## 8.1: AI in Education: An Overview

AI in education refers to using artificial intelligence technologies to enhance teaching and learning. These technologies can include machine learning algorithms, natural language processing, and intelligent tutoring systems. AI can personalize learning, facilitate assessment and feedback, and provide interactive and immersive learning experiences.

## 8.2: AI for Prompt Design

AI can assist in designing effective prompts by analyzing large volumes of educational data to identify patterns, trends, and insights. For example, AI algorithms can analyze students' responses to past prompts to determine which types of features of prompts are more effective in eliciting desired learning outcomes.

AI can also analyze subject-specific content, such as textbooks or research articles, to generate potential prompt questions or scenarios. Moreover, AI can personalize prompt design by considering individual students' learning styles, preferences, and performance data.

## 8.3: AI for Prompt Deployment

AI-powered platforms can facilitate prompt deployment in various ways. They can schedule and deliver prompts at optimal times or intervals based on pedagogical principles or individual learners' patterns and preferences. They can adapt prompts' sequence, complexity, or focus based on learners' progress or feedback.

AI can also facilitate interactive and adaptive prompts. For instance, chatbots can deliver Socratic prompts, adapting their questions or hints based on the learners' responses. Virtual tutors can guide learners through complex scenarios or problems, providing prompts and feedback as needed.

## 8.4: AI for Prompt Evaluation

AI can automate the evaluation of closed-ended prompts, providing instant grading and feedback. More advanced AI systems can even assess open-ended responses, using natural language processing and machine learning algorithms to analyze the responses' content, style, and quality.

AI can also analyze patterns and trends in the assessment data, identifying common issues, knowledge gaps, or learning trajectories. These insights can guide prompt refinement, curriculum development, and individualized learning support.

## 8.5: AI for Learner Engagement and Support

AI can enhance learner engagement and support in response to prompts. For example, AI-powered discussion forums can facilitate meaningful conversations around open-ended or reflective prompts, using algorithms to moderate the discussion, highlight insightful contributions, or suggest further prompts or resources.

AI-powered support systems can assist learners in responding to prompts, offering hints, resources, or scaffolding as needed. These systems can also monitor learners' progress and well-being, providing timely prompts for self-reflection, self-care, or seeking help.

## 8.6: Ethical and Practical Considerations

While AI offers exciting possibilities for prompt engineering, it also presents ethical and practical considerations. These include data privacy and security, fairness, bias in AI algorithms, the risk of over-reliance on AI, and human oversight and judgment. As we integrate AI into prompt engineering, we need to navigate these considerations responsibly, guided by ethical principles, pedagogical wisdom, and the best interests of our learners.

In conclusion, advanced AI technologies can significantly enhance prompt medical and nursing education engineering, offering new possibilities for personalization, interactivity, and data-driven insights. As we progress in this digital era, staying informed and critical about these technologies and leveraging their benefits while mitigating their challenges is crucial. In doing so, we can continue to advance our mission of fostering excellence in medical and nursing education, preparing our learners for a future where technology and healthcare are increasingly intertwined.

## 8.7: Questions

1. Discuss a recent research study or innovation in prompt engineering that could significantly impact medical and nursing education.
2. Reflect on how current research in cognitive science can inform the design of effective prompts.
3. How can innovations in AI and machine learning contribute to the evolution of prompt engineering?
4. Consider a research study on the use of prompts in medical education. How could its findings be applied to enhance nursing education?
5. Discuss the potential of virtual and augmented reality technologies in delivering immersive and interactive prompts.
6. Reflect on how research and innovations in prompt engineering can address the challenges of personalized and adaptive learning.
7. Evaluate the ethical considerations and potential unintended consequences of AI in prompt engineering.

8. Imagine the future of medical and nursing education with continuous advancements in prompt engineering. What opportunities and challenges do you foresee?
9. Discuss how research findings in prompt engineering can be translated into practical strategies for educators.
10. Reflect on the importance of evidence-based practice in designing and implementing prompts.

## 8.8: Key Takeaways

1. **AI in Education, An Overview:** AI technologies such as machine learning algorithms, natural language processing, and intelligent tutoring systems can enhance teaching and learning by personalizing education and facilitating assessment and feedback.
2. **AI for Prompt Design:** AI can assist in designing effective prompts by analyzing educational data to identify patterns, trends, and insights. It can analyze subject-specific content to generate potential prompt questions or scenarios and personalize prompt design based on individual students' learning styles and performance data.
3. **AI for Prompt Deployment:** AI-powered platforms can facilitate prompt deployment by scheduling and delivering prompts based on pedagogical principles or individual learners' patterns. They can also adapt the sequence and complexity of prompts based on learners' progress or feedback and facilitate interactive prompts.
4. **AI for Prompt Evaluation:** AI can automate the evaluation of closed-ended prompts and assess open-ended responses using advanced algorithms. It can analyze patterns in the assessment data to guide prompt refinement and curriculum development.
5. **AI for Learner Engagement and Support:** AI can enhance learner engagement and support in response to prompts. AI-powered forums can facilitate meaningful conversations around prompts, while AI-powered support systems can assist learners in responding to prompts and monitor learners' progress and well-being.

6. **Ethical and Practical Considerations:** As we integrate AI into prompt engineering, we need to consider data privacy and security, fairness and bias in AI algorithms, the risk of over-reliance on AI, and the need for human oversight and judgment.

## 8.9: Example Prompts for Chapter 8

1. AI for Prompt Design
    a. How can AI be used to analyze students' responses to past prompts to determine which types of features of prompts are more effective in eliciting desired learning outcomes?
    b. How can AI be used to analyze subject-specific content, such as textbooks or research articles, to generate potential prompt questions or scenarios?
    c. How can AI personalize prompt design by considering individual students' learning styles, preferences, and performance data?
2. AI for Prompt Deployment
    a. How can AI-powered platforms schedule and deliver prompts based on pedagogical principles or individual learners' patterns and preferences at optimal times or intervals?
    b. How can AI-powered platforms adapt prompts' sequence, complexity, or focus based on learners' progress or feedback?
    c. How can AI facilitate interactive and adaptive prompts, such as chatbots that deliver Socratic prompts or virtual tutors that guide learners through complex scenarios or problems?
3. AI for Prompt Evaluation
    a. How can AI automate the evaluation of closed-ended prompts, providing instant grading and feedback?
    b. How can more advanced AI systems be used to assess open-ended responses, using natural language processing and machine learning algorithms to analyze the responses' content, style, and quality?

  c. How can AI be used to analyze patterns and trends in the assessment data, identifying common issues, knowledge gaps, or learning trajectories?

4. AI for Learner Engagement and Support
    a. How can AI-powered discussion forums facilitate meaningful conversations around open-ended or reflective prompts, using algorithms to moderate the discussion, highlight insightful contributions, or suggest further prompts or resources?
    b. How can AI-powered support systems assist learners in responding to prompts, offering hints, resources, or scaffolding as needed?
    c. How can AI-support systems monitor learners' progress and well-being, providing timely prompts for self-reflection, self-care, or seeking help?

5. Ethical and Practical Considerations
    a. What are the ethical considerations for using AI for prompt engineering?
    b. What are the practical considerations for using AI for prompt engineering?
    c. How can we navigate these considerations responsibly, guided by ethical principles, pedagogical wisdom, and the best interests of our learners?

# Chapter 9: The Future of Prompt Engineering in Medical and Nursing Education

Prompt engineering has revolutionized how we teach and learn in medicine and nursing. As we look to the future, imagining how these techniques might evolve and adapt is essential. This chapter will speculate on the future of prompt engineering, considering emerging trends in technology, pedagogy, and healthcare.

## 9.1: A More Personalized Learning Journey

The future of prompt engineering may be deeply personalized, with prompts crafted and adapted to each learner's unique needs, preferences, and progress. Advanced AI technologies and learning analytics could enable real-time personalization of prompts, offering a truly individualized learning journey. Such personalization could help maximize each learner's potential and foster a more inclusive and learner-centered education system.

## 9.2: Greater Integration of Technology

We can expect to see an even greater technology integration into prompt engineering. Virtual, augmented, and mixed-reality platforms could provide immersive environments for role-play or scenario-based prompts. Haptic technologies could simulate tactile experiences, enhancing such prompts' realism and educational value.

Meanwhile, AI-powered intelligent tutoring systems could provide adaptive prompts and feedback, facilitating self-directed learning and continuous assessment. Brain-computer interfaces or neurofeedback technologies might enable prompts and feedback that respond to learners' cognitive or emotional states.

## 9.3: Expanding Roles and Skills

As prompt engineering evolves, so will educators' and learners' roles and skills. Educators must become adept at designing, deploying, and evaluating prompts in various formats and contexts, including digital and AI-powered platforms. They will need to stay attuned to emerging research, technologies, and pedagogical strategies and to navigate the ethical and practical issues of technology integration.

On the other hand, learners will need to become adept at responding to a broader range of prompts, reflecting on their responses, and using prompts and feedback for self-directed learning. They will also need to develop digital literacy, critical thinking, and metacognitive skills to navigate increasingly complex and technology-rich learning environments.

## 9.4: The Impact of Healthcare Trends

Trends in healthcare will also shape the future of prompt engineering. For instance, the increasing emphasis on patient-centered care, shared decision-making, and interdisciplinary teamwork could inspire new prompts that foster these competencies.

Similarly, healthcare's growing complexity and uncertainty could lead to more prompts that enhance problem-solving, critical thinking, and resilience. The surge of telemedicine, digital health, and precision medicine could require prompts that familiarize learners with these technologies and their implications.

## 9.5: Lifelong Learning and Professional Development

Prompt engineering could play a crucial role in lifelong learning and professional development, which are increasingly recognized as essential for healthcare professionals in the rapidly evolving medical field. Prompts could facilitate continuous self-assessment, reflection, and learning, keeping professionals updated and adaptable.

## 9.6: Research and Innovation

Lastly, we expect continued research and innovation in prompt engineering. This research could explore the effectiveness of various prompt types, strategies, and technologies, the impact of prompt engineering on learning outcomes and educational equity, and the best practices for prompt design, deployment, and evaluation.

In conclusion, the future of prompt engineering in medical and nursing education is full of exciting possibilities and challenges. As we continue this journey, let's remain committed to our core mission: fostering an enriching, empowering, and transformative learning experience that prepares our learners for a future that we can only imagine.

## 9.7: Questions

1. Reflect on how prompt engineering might evolve in the next decade in response to advancements in technology and cognitive science.
2. Discuss how future AI and data analytics trends could enhance personalization in prompt engineering.
3. How can future innovations in prompt engineering support the development of lifelong learning skills among medical and nursing professionals?
4. Consider a future scenario where VR is widely used in medical and nursing education. How might prompts be integrated into VR learning experiences?
5. Discuss how prompt engineering could contribute to tackling future challenges in healthcare education, such as the increasing complexity of patient care and the rapid pace of medical advancements.
6. Reflect on the future skills and competencies that medical and nursing educators might need to utilize prompt engineering effectively.
7. How can future trends in prompt engineering facilitate interprofessional and global collaboration in medical and nursing education?
8. Evaluate the potential ethical and societal implications of AI-powered prompt engineering in the future.
9. Imagine you're an educator in 2033. How would you utilize prompts to enhance learning in a mixed-reality classroom?

10. Reflect on your role as a lifelong learner in the evolving landscape of prompt engineering. How will you continue to adapt and grow?

## 9.9: Key Takeaways

1. **A More Personalized Learning Journey:** The future of prompt engineering may offer a deeply personalized learning experience thanks to advanced AI technologies and learning analytics.
2. **Greater Integration of Technology:** Technological advances such as virtual/augmented/mixed reality, haptic technologies, AI-powered tutoring systems, and neurofeedback technologies could all contribute to the evolution of prompt engineering.
3. **Expanding Roles and Skills:** Educators and learners must adapt as prompt engineering evolves. Educators will need to become proficient in new technologies and pedagogical strategies, while learners will need to develop a range of skills, including digital literacy, critical thinking, and metacognition.
4. **The Impact of Healthcare Trends:** Trends in healthcare, such as patient-centered care, shared decision-making, interdisciplinary teamwork, and the rise of telemedicine and digital health, will shape the future of prompt engineering.
5. **Lifelong Learning and Professional Development:** Prompt engineering could play a significant role in continuous learning and professional development for healthcare professionals, facilitating self-assessment, reflection, and adaptability.
6. **Research and Innovation:** Continued research and innovation in prompt engineering will contribute to a deeper understanding of the effectiveness of different prompt types, strategies, and technologies and to the development of best practices.

# 9.10: Example Prompts for Chapter 9

1. A More Personalized Learning Journey
   a. How can we use AI to personalize prompts for each learner, considering their unique needs, preferences, and progress?
   b. What are the ethical considerations of personalizing prompts?
   c. How can we ensure that personalization does not lead to discrimination or exclusion?
2. Greater Integration of Technology
   a. How can we use virtual, augmented, and mixed reality to create immersive and interactive learning experiences?
   b. How can haptic technologies simulate tactile experiences in virtual learning environments?
   c. How can AI-powered intelligent tutoring systems provide adaptive prompts and feedback?
3. Expanding Roles and Skills
   a. What new roles and skills will educators need to develop to use prompt engineering in their teaching effectively?
   b. What new roles and skills will learners need to develop to respond to prompts and use feedback for self-directed learning effectively?
   c. How can we support educators and learners in developing these new skills?
4. The Impact of Healthcare Trends
   a. How can prompt engineering foster patient-centered care, shared decision-making, and interdisciplinary teamwork?
   b. How can prompt engineering enhance healthcare professionals' problem-solving, critical thinking, and resilience?
   c. How can prompt engineering be used to familiarize learners with telemedicine, digital health, and precision medicine?
5. Lifelong Learning and Professional Development

a. How can prompt engineering be used to support lifelong learning and professional development for healthcare professionals?
b. How can we ensure prompt engineering is accessible and affordable for all learners?
c. How can we evaluate the effectiveness of prompt engineering for lifelong learning and professional development?

# Chapter 10: Strategies for Implementing Prompt Engineering in Your Teaching Practice

Having explored the theory and techniques of prompt engineering, it's time to apply these insights to your teaching practice. This final chapter will provide practical strategies for implementing prompt engineering, catering to various contexts and constraints.

## 10.1: Start with a Clear Learning Objective

Every prompt should start with a clear learning objective. The objective serves as a guide for designing the prompt and evaluating its effectiveness. Consider the knowledge, skills, or attitudes you want the learners to develop and how the prompt can facilitate that development. Consistently articulate the learning objective to your students, helping them understand the purpose of the prompt and focus their learning efforts.

## 10.2: Tailor Prompts to Your Learners

Understanding your learners is critical to designing effective prompts. Consider your learners' prior knowledge, skills, learning styles, and motivations. For instance, beginners might benefit from structured, scaffolded prompts, while advanced learners prefer challenging, open-ended prompts. Personalize the prompts as much as possible, catering to individual learners' needs and progress.

## 10.3: Experiment with Different Types of Prompts

Avoid sticking to one type of prompt; variety stimulates learning. Experiment with different types of prompts: closed-ended, open-ended, reflective, Socratic, and role-play prompts. Also, experiment with different formats or modalities, such as verbal, written, visual, or digital prompts. Encourage students to experiment with their responses, exploring different perspectives, approaches, or formats.

## 10.4: Integrate Prompts into Various Learning Activities

Prompts can be integrated into various learning activities: lectures, discussions, case studies, simulations, lab work, fieldwork, group projects, independent studies, etc. Think creatively about how prompts can enhance these activities, facilitating active engagement, deep thinking, meaningful interaction, or constructive feedback.

## 10.5: Provide Guidance and Support

Always provide guidance and support for responding to prompts. Clarify the expectations, provide examples or models, offer resources or scaffolding, and be available for questions or help. Create a supportive learning environment where students feel comfortable taking risks, making mistakes, and learning from them.

## 10.6: Elicit and Respond to Feedback

Elicit feedback from students about the prompts: Was the prompt clear? Did it facilitate learning? Was it too easy or difficult? What suggestions do they have for future prompts? Respond to the feedback constructively, refining your prompts and teaching practices accordingly.

## 10.7: Reflect on and Refine Your Prompt Engineering Skills

Reflect on your prompt engineering skills: What worked well? What challenges did you face? What did you learn? How can you improve? Seek professional development opportunities, such as workshops, courses, or peer coaching, to further refine your skills. Stay updated with the latest research and innovations in prompt engineering, medical, and nursing education.

## 10.8: Collaborate with Peers and Experts

Collaborate with peers and experts in your field or educational sciences. Share your experiences, ideas, and resources, learn from successes and challenges, and inspire each other to innovate and improve. Collaborations can also facilitate interdisciplinary prompts, broadening learners' perspectives and fostering teamwork skills.

## 10.9: Navigate the Challenges and Opportunities of Technology

If you use digital or AI-powered platforms for prompt engineering, navigate their challenges and opportunities responsibly. Ensure data privacy and security, maintain human oversight and judgment, and exploit the technology's unique features for enhancing learning.

In conclusion, implementing prompt engineering in your teaching practice is a journey of exploration, creativity, and growth. It requires patience, resilience, and a lifelong learning mindset. But the rewards are worth it: enhanced learning outcomes, a more engaging and empowering learning experience, and the joy of witnessing your students grow and succeed. As you embark on this journey, remember that you're not alone: You're part of a vibrant community of educators, researchers, and innovators, all striving to advance the noble mission of medical and nursing education.

## 10.10: Questions

1. Discuss the importance of the human touch in prompt engineering in medical and nursing education.
2. Reflect on an instance where an educator's empathy, creativity, or wisdom enhanced a prompt's impact.
3. How can prompts foster humanistic qualities such as empathy and professionalism among medical and nursing students?

4. Consider a sensitive topic in healthcare, such as end-of-life care or mental health. How could a well-crafted prompt facilitate a compassionate and nuanced understanding of this topic?
5. Discuss the role of educators in guiding AI-powered prompt engineering to ensure the preservation of humanistic values.
6. Reflect on how the human touch in prompt engineering can support the emotional well-being of learners, particularly in challenging fields such as palliative care or emergency medicine.
7. Evaluate the balance between technology and human interaction in prompt engineering. How can we leverage technology without losing the human touch?
8. Consider a complex ethical dilemma in healthcare. Design reflective prompts encouraging students to explore their values and emotions as future healthcare providers.
9. Discuss how human-centered prompt engineering can foster a culture of respect, empathy, and inclusivity in healthcare.
10. Reflect on your personal growth and development in medical or nursing education. How have prompts contributed to shaping not only your professional skills but also your personal qualities?

## 10.11: Key Takeaways

1. **Start with a Clear Learning Objective:** Every prompt should be based on a well-defined learning objective. Clearly articulating the objective to students helps guide their learning efforts.
2. **Tailor Prompts to Your Learners:** Understanding your learners' backgrounds, abilities, and motivations is critical for designing effective prompts. Personalization of prompts can significantly enhance learning outcomes.
3. **Experiment with Different Types of Prompts:** Variety in prompt type and format can help stimulate student interest and engagement. Encourage students to experiment with their responses.

4. **Integrate Prompts into Various Learning Activities:** Prompts can be incorporated into various learning activities, including lectures, discussions, and group projects, to facilitate active engagement and deep thinking.
5. **Provide Guidance and Support:** Clear expectations, resources, and scaffolding can help students respond to prompts effectively. A supportive learning environment encourages risk-taking and learning from mistakes.
6. **Elicit and Respond to Feedback:** Student feedback about the prompts is invaluable for refining your teaching practice. A constructive response to feedback shows students their opinions are valued.
7. **Reflect on and Refine Your Prompt Engineering Skills:** Regular self-reflection can help identify areas of improvement in your prompt engineering skills. Seek professional development opportunities to refine your abilities further.
8. **Collaborate with Peers and Experts:** Collaboration with colleagues and experts can provide fresh perspectives, shared resources, and inspiration. It can also foster interdisciplinary learning.
9. **Navigate the Challenges and Opportunities of Technology:** If using digital or AI-powered platforms, ensure data privacy and security, maintain human oversight, and maximize the potential of technology for learning enhancement.

## 10.12: Example Prompts for Chapter 10

1. Start with a Clear Learning Objective
    a. What do you want your learners to learn from this prompt?
    b. What knowledge, skills, or attitudes do you want them to develop?
    c. How will you know if they have achieved the learning objective?
2. Tailor Prompts to Your Learners
    a. What are your learners' prior knowledge, skills, and learning styles?
    b. What are their interests and motivations?
    c. How can you tailor the prompt to meet their individual needs?
3. Experiment with Different Types of Prompts

a. What type of prompt will best engage your learners and facilitate learning?
   b. Consider closed-ended, open-ended, reflective, Socratic, and role-play prompts.
   c. Also, experiment with different formats or modalities, such as verbal, written, visual, or digital prompts.
4. Integrate Prompts into Various Learning Activities
   a. How can you integrate prompts into your lectures, discussions, case studies, simulations, lab work, fieldwork, group projects, or independent studies?
   b. Think creatively about how prompts can enhance these activities, facilitating active engagement, deep thinking, meaningful interaction, or constructive feedback.
5. Provide Guidance and Support
   a. How will you provide guidance and support for responding to prompts?
   b. Clarify the expectations, provide examples or models, offer resources or scaffolding, and be available for questions or help.
   c. Create a supportive learning environment where students feel comfortable taking risks, making mistakes, and learning from them.

# Chapter 11: The Role of Diversity, Equity, and Inclusion in Prompt Engineering

Diversity, equity, and inclusion (DEI) are essential principles in the practice of medicine. DEI principles ensure that all individuals are treated fairly and compassionately regardless of their background or identity.

## 11.1: Understanding Diversity, Equity, and Inclusion

Diversity refers to the presence of heterogeneity in groups. Essential areas of diversity include differences in race, ethnicity, gender, sexual orientation, socioeconomic status, ability, nationality, and more. Equity, on the other hand, is about ensuring fair treatment. This includes open access to resources and opportunities. Barriers to access must be addressed. Inclusion goes further to actively involve and value all people, celebrating diversity and promoting a sense of belonging.

## 11.2: The Importance of DEI in Prompt Engineering

DEI is crucial in prompt engineering for several reasons. Firstly, diverse and inclusive prompts can validate and value diverse identities, perspectives, and experiences, promoting a sense of belonging and respect among all learners. Secondly, they can expose learners to a broader range of perspectives and contexts, fostering cultural competence and interdisciplinary thinking. Finally, equitable prompts can ensure that all learners, including those from marginalized or disadvantaged backgrounds, have fair opportunities to engage, learn, and succeed.

## 11.3: Strategies for Diverse and Inclusive Prompts

Creating diverse and inclusive prompts involves several strategies. Consider using prompts representing various identities, perspectives, or contexts. This might involve using case studies from different cultural settings or including prompts that challenge gender or racial stereotypes.

Secondly, ensure that your prompts do not inadvertently exclude or marginalize certain groups. Avoid using jargon, stereotypes, or assumptions that might alienate some learners. Also, consider the accessibility of your prompts, ensuring they can be understood and responded to by learners with different abilities, language proficiencies, or technological access.

Thirdly, consider using prompts that foster inclusive dialogues or interactions. For example, you might use prompts to encourage learners to share their experiences or perspectives. These prompts facilitate respectful debates on controversial issues or prompts that stimulate collaborative problem-solving.

## 11.4: Strategies for Equitable Prompts

Creating equitable prompts also involves several strategies. Firstly, provide clear, detailed, and accessible instructions for all prompts. This ensures that all learners, regardless of their background, have a good chance of understanding and responding to the prompt.

Secondly, differentiate your prompts according to learners' diverse needs or levels. You might provide additional scaffolding for learners who need more support or extra challenges for learners who need more stimulation. AI technologies can be particularly helpful in personalizing prompts for each learner's pace and level.

Thirdly, provide equal opportunities for learners to respond to prompts. Ensure that all learners have the time, resources, and support to engage with the prompts. Also, create a safe and respectful learning environment where all learners feel comfortable expressing their thoughts or asking questions.

## 11.5: Evaluating DEI in Prompt Engineering

Assessing the impact of your DEI strategies is an integral part of the process. Gather feedback from your students regarding their experiences with the prompts, whether they felt included, respected, and challenged. Use this feedback to refine your strategies, ensuring they effectively promote diversity, equity, and inclusion.

## 11.6: Continuous Learning and Growth in DEI

Like prompt engineering, promoting DEI is a journey of continuous learning and growth. Stay updated with DEI's latest research and best practices, and seek professional development opportunities in this area. Don't be afraid to make mistakes or face challenges; they can be valuable learning experiences. Remember, your efforts to promote DEI can profoundly impact your learners, enhancing their learning outcomes and fostering their personal growth, interpersonal skills, and social responsibility.

In conclusion, integrating DEI principles into prompt engineering is a powerful strategy to enhance medical and nursing education's quality, relevance, and impact. It helps us cultivate diversity, equity, and inclusiveness. We are enriched by the multiple perspectives and experiences that each member brings. It is an investment in our learners, professions, and society and a step towards a more equitable and inclusive future.

## 11.7: Questions

1. Reflect on a scenario where you have encountered or observed inequity or exclusion in a learning environment. How could prompt engineering be used to address these issues?

2. Identify and analyze a prompt you've used in your teaching that might have unintentionally marginalized or excluded some learners. How could you revise this prompt to make it more inclusive and equitable?
3. Imagine you are teaching a diverse class of medical or nursing students. Design a prompt to engage and validate all their diverse identities, perspectives, and experiences.
4. Propose a strategy for differentiating your prompts according to your learners' diverse needs, backgrounds, and abilities. Provide examples of how you might adapt a specific prompt for different learners.
5. Discuss how prompt engineering can foster cultural competence among medical or nursing students. Design a prompt that can facilitate such competence.
6. Describe how you might collect and analyze feedback from your students about the diversity, equity, and inclusivity of your prompts. What specific questions might you ask? What steps might you take based on their feedback?
7. Reflect on a mistake or challenge in integrating DEI principles into your prompt engineering. What did you learn from this experience, and how might you address this challenge in the future?
8. Identify a recent research study or innovation related to DEI in medical or nursing education. Discuss its implications for prompt engineering, providing specific examples of how it might inform your prompts.
9. Imagine you are collaborating with a fellow educator who is new to DEI principles. How would you explain the importance of DEI in prompt engineering, and what advice would you give them for integrating DEI into their prompts?
10. Reflect on your journey of promoting DEI through prompt engineering. What accomplishments are you proud of, and what goals do you have for the future?

## 11.8: Key Takeaways

1. Diversity, equity, and inclusion (DEI) are crucial principles that ensure fair treatment, a sense of value, and equal opportunities for all individuals, regardless of their backgrounds or identities.
2. DEI is essential in prompt engineering as it validates diverse identities, perspectives, and experiences, fosters cultural competence, and ensures equal opportunities for all learners.
3. To create diverse and inclusive prompts, one should represent various identities and perspectives, avoid alienating or marginalizing certain groups, and facilitate inclusive dialogues or interactions.
4. Equitable prompts require clear, detailed, and accessible instructions, differentiation according to learners' diverse needs, and equal opportunities for learners to engage with the prompts.
5. Evaluating DEI in prompt engineering involves gathering feedback from learners about their experiences and using it to refine DEI strategies.
6. DEI promotion is a continuous learning and growth journey involving staying updated with the latest research, seeking professional development, and learning from mistakes.
7. Integrating DEI principles into prompt engineering can significantly enhance medical and nursing education's quality, relevance, and impact.

## 11.9: Example Prompts for Chapter 11

1. Diversity
    a. How do you ensure that your prompts represent a diversity of identities, perspectives, and experiences?
    b. How can you avoid using jargon, stereotypes, or assumptions that might alienate some learners?

c. How can your prompts be accessible to learners with different abilities, language proficiencies, or technological access?
2. Equity
   a. How can you provide clear, detailed, and accessible prompt instructions?
   b. How can you differentiate your prompts according to learners' needs or levels?
   c. How can you provide equal opportunities for learners to respond to prompts?
3. Inclusion
   a. How can you create prompts that foster inclusive dialogues or interactions?
   b. How can you create a safe and respectful learning environment where all learners feel comfortable expressing their thoughts or asking questions?
   c. How can you gather feedback from your students regarding their experiences with the prompts, whether they felt included, respected, and challenged?
4. Continuous Learning and Growth
   a. How can you stay updated with the latest research and best practices in DEI?
   b. How can you seek professional development opportunities in this area?
   c. How can you overcome challenges and mistakes in your efforts to promote DEI?
5. Impact
   a. How can you assess the impact of your DEI strategies?
   b. How can you use this feedback to refine your strategies continuously?
   c. How can you ensure that your efforts to promote DEI profoundly impact your learners?

# Conclusion: Harnessing the Power of Prompt Engineering in Medical and Nursing Education

Throughout this guide, we have traversed the landscape of prompt engineering, exploring its foundational principles, diverse techniques, and promising advancements. We've delved into the cognitive science underpinnings of effective prompts, dissected the anatomy of various prompts, and envisioned the future of prompt engineering infused with advanced AI technologies. We've seen how prompt engineering can enhance teaching and learning in medical and nursing education, fostering more profound understanding, critical thinking, reflection, and engagement.

Prompt engineering is more than just an instructional tool or strategy. It embodies an educational philosophy that values active learning, critical thinking, and self-directed learning. It celebrates the diversity and potential of each learner, advocating for personalized, inclusive, and empowering education. It strives to bridge the gap between theory and practice, knowledge and skills, and the classroom and the clinic, preparing our learners for the complexities and uncertainties of healthcare.

As we conclude this journey, we invite you – whether you are a student, teacher, or educational leader – to embrace the power and potential of prompt engineering. Here are some key takeaways to inspire your journey:

1. **Start with the Learner:** Every prompt should start and end with the learner. What do they already know? What do they need to learn? How do they learn best? Use prompts to meet learners where they are and guide them toward where they need to be.
2. **Diversify Your Prompts:** Use various prompts to stimulate various cognitive processes, cater to different learning styles, and keep learning engaging and challenging. Remember, there's no one-size-fits-all prompt.

3. **Embrace the Socratic Method:** Make the most of Socratic prompts to foster critical thinking, reflection, and dialogue. Encourage learners to question their assumptions, consider multiple perspectives, and construct their understanding.
4. **Harness the Power of Technology:** Leverage digital and AI technologies to augment your prompt engineering. Use these technologies to personalize prompts, facilitate interactive and immersive learning experiences, and derive data-driven insights.
5. **Iterate and Improve:** Prompt engineering is an art and science that requires ongoing practice, reflection, and refinement. Don't be afraid to experiment, make mistakes, and learn from them. Seek feedback from your learners, collaborate with your peers, and stay updated with the latest research and innovations.
6. **Keep the Human Touch:** Despite the increasing integration of technology, the human touch remains crucial in prompt engineering. The educator's empathy, creativity, and wisdom breathe life into the prompts, sparking the learners' curiosity, passion, and growth. The learners' voices, responses, and transformations bring the prompts to fruition, enriching the collective learning journey.
7. **Change the World, One Prompt at a Time:** Every prompt is a small step towards our big dreams: a world where every learner thrives, every patient is cared for, and every health challenge is tackled. We are shaping minds, touching hearts, and changing lives with every thoughtful prompt.

As we look to the future, we are excited about the potential of prompt engineering to revolutionize medical and nursing education and beyond. We are inspired by the community of educators, learners, researchers, and innovators pioneering this field, turning prompts into promises, questions into quests, and challenges into changes.

And so, we urge you: Take the plunge. Experiment with prompts. Revel in the thrill of discovery, the beauty of thinking, and the joy of learning. Join us on this incredible journey of prompt engineering – a journey that promises to reshape education, empower learners, and transform healthcare, one prompt at a time.

# Appendix 1: Annotated Bibliography

Prompt Engineering with ChatGPT. *ArXiv*. https://doi.org/10.48550/arxiv.2302.11382

*This paper presents a prompt pattern catalog to help with prompt engineering. The proposed catalog contains a set of pre-defined prompt patterns that can be used to improve the performance of LLMs on various tasks. The catalog is evaluated on a set of tasks and shown to improve the performance of LLMs on these tasks significantly.*

4. Zhou, Y., Muresanu, A. I., Han, Z., Paster, K., Pitis, S., Chan, H., & Ba, J. (2022). Large Language Models Are Human-Level Prompt Engineers. *ArXiv*. https://doi.org/10.48550/arxiv.2211.01910

*Large language models (LLMs) can generate text, translate languages, write different kinds of creative content, and answer your questions in an informative way. However, the performance of LLMs on these tasks depends significantly on the quality of the prompt used to steer the model. In this paper, the authors propose a new method for automatically generating prompts that can be used to improve the performance of LLMs on various tasks. The proposed method, called Automatic Prompt Engineer (APE), is evaluated on various tasks and shown to improve the performance of LLMs on these tasks significantly.*

5. Wang, J., Shi, E., Yu, S., Wu, Z., Ma, C., Dai, H., Yang, Q., Kang, Y., Wu, J., Hu, H., Yue, C., Zhang, H., Liu, Y., Li, X., Ge, B., Zhu, D., Yuan, Y., Shen, D., Liu, T., & Zhang, S. (2023). Prompt Engineering for Healthcare: Methodologies and Applications. *ArXiv*. https://doi.org/10.48550/arxiv.2304.14670

*Prompt engineering is a technique that can be used to improve the performance of large language models (LLMs) on a variety of tasks in healthcare, such as question-answering, summarization, and translation. The authors of this paper review the latest advances in prompt engineering for healthcare and discuss the potential applications of prompt engineering in this field. They conclude that prompt*

*Prompt engineering is crafting instructions that tell large language models (LLMs) what to do. By carefully crafting prompts, engineers can get LLMs to generate text, translate languages, write different kinds of creative content, and answer your questions in an informative way. Prompt engineering is a powerful new tool that has the potential to revolutionize the way we interact with AI.*

9. Bal Ram, & Pratima Verma. (2023). Artificial intelligence AI-based Chatbot study of ChatGPT, Google AI Bard, and Baidu AI. *World Journal of Advanced Engineering Technology and Sciences*, *8*(1), 258–261. https://doi.org/10.30574/wjaets.2023.8.1.0045

*This study compares three large language models (LLMs) for chatbot applications: ChatGPT, Google AI Bard, and Baidu AI. The study found that all three models could generate human-like text and engage in conversations, but Google AI Bard was the most accurate and consistent. The study also found that all three models could translate languages, write different kinds of creative content, and answer questions informatively. However, Google AI Bard was the most comprehensive and informative. The study found that Google AI Bard is the most promising LLM for chatbot applications. Caution: This study used ChatGPT-3.5, not the newer ChatGPT-4, which is much more powerful.*

# Appendix 2: Resources

1. *Learn Prompting: Your Guide to Communicating with AI*. (n.d.). Retrieved May 21, 2023, from https://learnprompting.org/
2. *Matt Wolfe - YouTube*. (n.d.). Retrieved May 21, 2023, from https://www.youtube.com/@mreflow
3. *Jason West - YouTube.* (n.d.) Retrieved May 21, 2023, from https://www.youtube.com/@JWestdigital
4. CondyOnChain - YouTube. (n.d.) Retrieved May 21, 2023, from https://www.youtube.com/@CodyOnChain

# Appendix 3: Example Prompts to Large Language Models Including ChatGPT and Google Bard

Try the following prompts on yourself without any assistance from the Internet, a textbook, or AI. Then enter the prompts into ChatGPT or Google Bard, engaging the AI with additional questions that arise.

## Anatomy and physiology

1. Embark on a captivating journey as a medical student exploring the remarkable intricacies of the cardiovascular system's normal physiology, unraveling the secrets of its anatomy and functions through engaging practical scenarios.
2. Join the expedition as a medical student, immersing yourself in the fascinating world of the cardiovascular system's normal physiology, unlocking the enigmas of its anatomy and functions through hands-on practical scenarios.
3. What is the primary cause of myocardial infarction?
4. Describe the anatomical structures involved in the pathogenesis of ischemic stroke.
5. How does the blockage of a coronary artery lead to the development of myocardial infarction?
6. Assume the role of a healthcare provider explaining the anatomy of a heart attack to a patient. How would you communicate the location and impact of the blocked blood vessel?
7. Compare and contrast the anatomical differences between an ischemic and a hemorrhagic stroke, focusing on the underlying mechanisms and affected brain regions.
8. Create an illustration showing the coronary arteries and their major branches, highlighting the areas most susceptible to myocardial infarction.

9. Investigate the role of atherosclerosis in developing both myocardial infarction and stroke. Summarize the anatomical changes that occur in the affected blood vessels.
10. Discuss the ethical considerations surrounding using thrombolytic therapy in treating acute myocardial infarction and ischemic stroke.
11. If a patient presents with sudden-onset neurological deficits, what anatomical and imaging studies would you perform to determine the type and location of the stroke?
12. Imagine a recent myocardial infarction patient who develops complications such as ventricular fibrillation. Discuss the anatomical basis for these complications and propose appropriate management strategies.

## Physiology

1. What is the primary defect in insulin secretion in type 1 diabetes?
2. Describe the physiological mechanisms underlying insulin resistance in type 2 diabetes.
3. How does chronic hyperglycemia contribute to the development of microvascular complications in diabetes?
4. Assume the role of a patient newly diagnosed with type 2 diabetes. Discuss how you would modify your lifestyle and daily routine to manage the disease effectively.
5. Compare and contrast the physiological differences between fasting plasma glucose and postprandial glucose levels in diabetes, emphasizing the underlying hormonal regulation.
6. Create an anatomical diagram illustrating the interaction between pancreatic beta cells, liver, skeletal muscle, and adipose tissue in glucose homeostasis.
7. Explore recent advancements in understanding the role of gut microbiota in the development and progression of diabetes. Summarize key findings and their implications.
8. Discuss the ethical considerations surrounding using new technologies, such as closed-loop insulin delivery systems (artificial pancreas), in managing diabetes.

9. If a patient presents with classic symptoms of diabetes, what physiological tests would you perform to confirm the diagnosis and differentiate between type 1 and type 2 diabetes?
10. Imagine a patient with poorly controlled diabetes who experiences frequent hypoglycemic episodes. Discuss the physiological reasons behind hypoglycemia and propose management strategies to prevent future episodes.
11. Imagine you are a glucose molecule on a journey through the body. Describe your path and the physiological processes in various tissues during normal glucose metabolism and diabetes.

## Biochemistry

1. What is the primary metabolic abnormality in gout?
2. Describe the process of purine metabolism and its relevance to the development of gout.
3. How does the accumulation of uric acid crystals in the joints contribute to the characteristic symptoms of gout?
4. Imagine you are explaining the mechanism of action of allopurinol, a medication commonly used in gout treatment, to a patient. How would you communicate its biochemistry and its impact on uric acid levels?
5. Compare and contrast the biochemical basis of primary and secondary gout, focusing on the underlying causes and contributing factors.
6. Create a diagram illustrating the steps in converting purines to uric acid, highlighting the enzymes involved and their regulation.
7. Investigate the role of dietary factors, such as purine-rich foods and fructose, in developing and exacerbating gout. Summarize your findings.
8. Discuss the potential ethical implications of using genetic testing to identify individuals at a higher risk of developing gout due to specific genetic variations.
9. If a patient presents with joint pain and swelling, how would you use laboratory tests to confirm a diagnosis of gout? Describe the relevant biochemical markers to consider.

10. Imagine a patient with gout who has not responded well to conventional treatments. Propose a novel therapeutic approach based on your understanding of the underlying biochemistry of gout.
11. Embark on a biochemical journey through the purine salvage pathway. Describe the key enzymes and molecules involved, highlighting their significance in gout development.
12. Imagine you are a uric acid crystal inside a joint affected by gout. Describe the biochemical interactions between the crystal and surrounding tissues and the resulting inflammatory response.

## Taking a Medical History

1. What is the purpose of taking a medical history?
2. Describe a comprehensive medical history's components and why each is important.
3. How does the order and structure of your questions during an interview impact the accuracy and completeness of the medical history obtained?
4. Assume the role of a healthcare provider conducting a medical interview. Practice asking open-ended questions to gather information about a patient's chief complaint.
5. Imagine you are a medical detective investigating a complex case. Create a list of essential questions you would ask during the medical history to gather crucial information.
6. Compare and contrast the approach to taking a medical history in a pediatric patient versus an elderly patient, highlighting each population's unique considerations and challenges.
7. Design a visual aid or flowchart that outlines the sequential steps and critical questions to ask when taking a medical history, considering both general and specialized focus areas.
8. If a patient presents with a specific complaint, such as chest pain or abdominal discomfort, list the targeted questions you would ask to gather pertinent information for an accurate diagnosis.

9. Discuss the ethical considerations and challenges when obtaining a medical history from a non-communicative patient with cognitive impairments or from a patient with a different cultural background than yours.
10. Explore the impact of patient-centered interviewing techniques on the quality of the medical history obtained and patient satisfaction. Summarize the findings from relevant studies.
11. Imagine you are a medical history explorer traveling through time. Select a specific historical period and describe the key differences and similarities in interviewing techniques compared to contemporary practices.
12. Reflect on the importance of active listening and empathy during a medical interview. Discuss how these skills contribute to building rapport, and trust, and gathering accurate patient information.

## Doing the Physical Exam

1. What is the purpose of a physical exam in medical practice?
2. Describe a comprehensive physical examination's general approach and sequence, highlighting the key components.
3. How does the examiner's knowledge of anatomy and physiology contribute to a thorough physical examination?
4. Assume the role of a healthcare provider conducting a physical exam. Practice explaining each step of the examination to a simulated patient.
5. Imagine you are an explorer in the human body. List the main landmarks and techniques you would use to assess the cardiovascular system during a physical exam.
6. Compare and contrast the physical examination approaches for a pediatric patient versus an adult patient, focusing on each population's unique considerations and adaptations.
7. Design a visual aid, such as a chart or diagram, that outlines the steps and key findings of a thorough physical examination of various body systems.

8. If a patient presents with a specific complaint, such as shortness of breath or abdominal pain, describe the targeted physical examination maneuvers you would perform to gather relevant clinical information.
9. Discuss the importance of cultural sensitivity and inclusion when conducting a physical examination. Reflect on the potential challenges and strategies for providing respectful care to individuals from diverse backgrounds.
10. Explore the role of technological advancements, such as point-of-care ultrasound or digital auscultation devices, in enhancing the accuracy and efficiency of physical examinations. Summarize key findings from relevant studies.
11. Imagine you are a detective investigating a puzzling case. List the specialized physical examination maneuvers and tests you would use to gather critical information for accurate diagnosis.
12. Reflect on the significance of thorough documentation during a physical exam. Discuss how accurate and detailed records contribute to effective communication, continuity of care, and patient safety.

## Medicolegal Issues

1. What is the primary purpose of medical malpractice laws?
2. Describe the key legal responsibilities and obligations of healthcare professionals in the delivery of patient care.
3. How does informed consent protect patients' rights and healthcare professionals' legal interests?
4. Assume the role of a physician encountering a challenging medicolegal scenario. Discuss the ethical and legal considerations that would influence your decision-making process.
5. Imagine you are a lawyer specializing in healthcare law. List the potential legal risks and liabilities that doctors and nurses should consider in their practice.
6. Compare and contrast the legal frameworks governing the practice of medicine in different countries, considering variations in liability, licensure, and scope of practice.

7. Design an infographic highlighting the steps healthcare professionals should take to mitigate medicolegal risks, such as maintaining accurate medical records, adhering to confidentiality, and practicing within their expertise.
8. If a healthcare professional faces a medicolegal complaint, outline the steps to address the situation effectively while minimizing legal risks.
9. Discuss the importance of diversity, equity, and inclusion in the context of medicolegal aspects of healthcare. Reflect on how considerations of cultural competency and patient autonomy impact legal obligations.
10. Explore recent landmark cases in the healthcare laws that have influenced the medicolegal landscape. Summarize the implications of these cases for healthcare professionals and patient care.
11. Imagine you are an advocate for patient rights. Develop a guidebook that educates patients on their legal rights and responsibilities when seeking medical care.
12. Reflect on professional liability insurance and its role in mitigating medicolegal risks for healthcare professionals. Discuss the benefits and challenges of obtaining and maintaining such insurance coverage.

## Avoiding Burnout

1. What are the common causes of burnout among doctors and nurses?
2. Describe the signs and symptoms of burnout and its impact on the well-being of healthcare professionals.
3. How does work-life balance contribute to preventing burnout among doctors and nurses?
4. Assume the role of a healthcare professional experiencing burnout. Discuss strategies to create a healthier work environment and reduce burnout.
5. Imagine you are a traveler exploring the world of self-care. List different self-care practices that doctors and nurses can incorporate daily to prevent burnout.
6. Compare and contrast the challenges and strategies for preventing burnout in different healthcare settings, such as hospitals, clinics, or community healthcare centers.

7. Design a poster promoting a healthy work environment and tips for preventing burnout, such as fostering social connections, setting boundaries, and practicing self-compassion.
8. If a healthcare professional is experiencing symptoms of burnout, outline the steps they should take to seek support and resources for addressing burnout effectively.
9. Discuss the intersection of diversity, equity, and inclusion with burnout prevention in healthcare. Reflect on the importance of addressing systemic factors contributing to burnout among marginalized healthcare professionals.
10. Explore recent studies on interventions and programs to prevent burnout among doctors and nurses. Summarize the essential findings and recommendations for burnout prevention.
11. Imagine you are a mentor guiding junior healthcare professionals on their journey to prevent burnout. Develop a roadmap that outlines strategies for maintaining well-being throughout their careers.
12. Reflect on resilience and its role in preventing burnout among doctors and nurses. Discuss how fostering resilience can contribute to improved coping mechanisms and overall well-being.

## Acute Cholecystitis

1. Is acute cholecystitis a life-threatening condition?
2. Describe the typical presentation and symptoms of acute cholecystitis, and discuss the potential complications if left untreated.
3. How does gallstone obstruction lead to the development of acute cholecystitis, and what are the consequences on the gallbladder?
4. Assume the role of a gallstone detective. Investigate the adventures of a gallstone as it navigates through the biliary tract, causing acute cholecystitis.
5. Imagine you are an explorer in the land of the gallbladder. Describe the breathtaking landscapes and hidden dangers that may contribute to acute cholecystitis.

6. Compare and contrast the clinical features and management of acute cholecystitis in young adults versus elderly patients, considering each group's unique challenges and considerations.
7. Create a whimsical cartoon strip illustrating the journey of a gallstone causing acute cholecystitis, highlighting the key steps and events leading to inflammation.
8. If a patient presents with suspected acute cholecystitis, describe the diagnostic tests and imaging modalities that can aid in confirming the diagnosis and assessing disease severity.
9. Discuss the impact of demographic factors, such as age, gender, or ethnicity, on acute cholecystitis's risk, presentation, and outcomes. Reflect on potential health disparities and equitable access to care.
10. Explore emerging technologies or futuristic interventions for managing acute cholecystitis, such as nanobots that can dissolve gallstones or virtual reality simulations for surgical training.
11. Imagine you are a time traveler visiting ancient civilizations. Investigate how acute cholecystitis was perceived, diagnosed, and treated in different historical periods.
12. Reflect on the concept of gallbladder resilience and its role in preventing acute cholecystitis. Discuss factors that can promote gallbladder health and reduce the risk of inflammation.

## Respiratory Diseases in Children

1. What is the most common respiratory disease in children?
2. Describe the signs and symptoms of respiratory diseases in children and how they differ from those in adults.
3. How does understanding the anatomy and physiology of the respiratory system help diagnose and manage respiratory diseases in children?
4. Assume the role of a pediatric respiratory therapist. Demonstrate how you would teach a child how to use an inhaler through role-playing properly.
5. Imagine you are a detective solving a mystery in the lungs. Describe the adventures and challenges you face while investigating respiratory diseases in children.

6. Discuss the importance of diversity, equity, and inclusion in the context of respiratory diseases in children. Reflect on the potential impact of socioeconomic factors on access to care and disease outcomes.
7. Explore the historical perspective of respiratory diseases in children. Discuss the evolution of medical knowledge, diagnostic tools, and treatments.
8. Imagine a futuristic world where respiratory diseases in children are eradicated. Describe the technological advancements and interventions that have made this possible.
9. How does exposure to environmental factors, such as air pollution or allergens, contribute to developing respiratory diseases in children?
10. Investigate the role of vaccinations in preventing respiratory diseases in children. Discuss their effectiveness and impact on public health.
11. Embark on a virtual reality adventure through the respiratory system of a child. Explore how infections spread and affect different lung structures.
12. Reflect on the psychosocial impact of respiratory diseases in children. Discuss their emotional and social challenges and strategies to support their overall well-being.

## Osteoporosis

1. What are the risk factors and preventive measures for osteoporosis?
2. How does osteoporosis affect the quality of life for older adults?
3. Using the Socratic method, explore the impact of gender inequality on the prevalence of osteoporosis in society.
4. Imagine you are a healthcare provider discussing the importance of exercise and nutrition in preventing osteoporosis with a sedentary patient. Role-play the conversation.
5. Embark on an adventure through time, visiting different historical periods to understand how osteoporosis was perceived and treated in each era.
6. How can healthcare institutions ensure diversity, equity, and inclusion in the diagnosis and treatment of osteoporosis?

7. Take a futuristic perspective and envision the potential advancements in technology that could revolutionize the diagnosis and treatment of osteoporosis in the next 50 years.
8. You encounter a patient with a fractured hip and suspect osteoporosis. Develop a clinical scenario to guide the diagnostic process and treatment options.
9. Discuss the ethical considerations surrounding the allocation of limited healthcare resources for the treatment of osteoporosis in elderly populations.
10. Compare and contrast the diagnostic methods and criteria for osteoporosis in men and women, highlighting any gender-specific considerations.
11. Conceptualize a novel intervention or treatment approach for osteoporosis that targets the underlying mechanisms of bone loss and promotes bone regeneration.
12. If you were as curious as a cat, what mystical powers would you associate with osteoporosis? How would those powers affect individuals who possess them?
13. In a world where a mysterious curse causes osteoporosis, explore the journey of a young healer who must unravel the ancient enchantments and discover a remedy to break the curse.
14. Imagine a hidden realm where ancient guardians protect the secrets of solid bones and everlasting vitality. Write a mythical tale of a chosen one who must undergo extraordinary trials to obtain the knowledge and power to prevent and heal osteoporosis in the mortal realm.

## Osteoarthritis of the knees

1. How would you differentiate between the symptoms of osteoarthritis and rheumatoid arthritis in the knee?
2. Imagine you are in the 19th century, and a patient comes to you complaining of knee pain. How would you diagnose and treat this patient's condition based on the available knowledge?
3. How can advances in biotechnology and genetic engineering contribute to better understanding and managing knee osteoarthritis in the future?

4. As a medical professional in a diverse community, how would you ensure equitable access to osteoarthritis care and culturally sensitive patient education?
5. In a rural town, you are the only medical professional available, and a patient walks in with severe knee pain. Based on your suspicions of osteoarthritis, how would you manage the situation, given limited resources?
6. What are the ethical implications of recommending knee replacement surgery for osteoarthritis, considering its high cost, potential complications, and quality of life for the patient?
7. You are presented with a patient complaining of knee pain and difficulty walking. After a physical examination and X-rays, you suspect osteoarthritis. How would you go about making a definitive diagnosis?
8. How do the pathophysiology and treatment strategies for knee osteoarthritis differ from those of the hip?
9. Given the prevalence of knee osteoarthritis, what would be the broader implications for society if a cure were discovered?
10. You've just discovered a diary from your cat detailing its observations of your behavior when your osteoarthritis of the knee flares up. What might it say?
11. Envision a scenario where you're not a doctor but a time-traveling healer with magical herbs that can relieve pain. How would you describe osteoarthritis of the knee and its treatment to a medieval villager?
12. You find an old, dust-covered medical book in a dimly lit room. As you wipe away the dust, the title is "The Secret Life of Knee Joints." The book claims to hold the mysteries of knee osteoarthritis. What do you hope to find inside?

## Gastroesophageal Reflux Disease

1. Is gastroesophageal reflux disease (GERD) more prevalent in any demographic group?
2. Explore the potential impacts of lifestyle and diet modifications on the management of GERD. What kinds of changes might be most effective?

3. Consider this: if stomach acid helps break down food in the stomach, then why does the backward flow of this acid into the esophagus cause harm in the case of GERD?
4. You're a specialist brought in to give a talk to primary care physicians about GERD. What key points would you like to address in your talk, particularly those that these physicians might overlook in their day-to-day practice?
5. You are a medical explorer journeying through the human digestive system. Describe your experience as you witness the processes leading up to GERD.
6. In your practice, you notice a trend of lower diagnosis rates for GERD among marginalized populations. What steps could you take to ensure more equitable healthcare provision in this context?
7. How has our understanding and treatment of GERD evolved over the centuries, and how has it shaped current medical practices?
8. Given the advancements in artificial intelligence and robotics, what could be the future of diagnosing and treating GERD?
9. You are presented with a patient experiencing chronic heartburn and regurgitation for the past six months. What steps would you take to diagnose and treat this patient?
10. Discuss the potential ethical issues involved in the long-term prescription of proton pump inhibitors (PPIs) for GERD patients.
11. You're faced with two patients with similar symptoms. One is diagnosed with GERD, and the other with a peptic ulcer. How would you differentiate between these conditions based on their symptoms and diagnostic tests?
12. The cat you live with starts acting differently around you when your GERD symptoms begin to flare up. What changes do you observe in its behavior?
13. You've come across an ancient scroll claiming to have the solution to "The Fire of the Drake," an old term for GERD. What could this mysterious cure be?
14. You're presented with a seemingly ordinary case of GERD. However, the patient doesn't respond to conventional treatment, and their condition remains unexplained. How would you approach this medical mystery?

# Acute Abdominal Pain

1. What are the most common causes of acute abdominal pain?
2. Discuss the different ways that acute abdominal pain can be managed.
3. What are the risk factors for acute abdominal pain? Can you think of any other risk factors that are not listed here?
4. You are a doctor seeing a patient experiencing acute abdominal pain. What do you do?
5. You are on a quest to find the cure for all diseases. Along the way, you come across a cave that is said to be filled with people suffering from acute abdominal pain. What do you do?
6. How does acute abdominal pain disproportionately affect specific populations? How can we ensure everyone can access the care they need for acute abdominal pain?
7. How has our understanding of acute abdominal pain changed over time?
8. What are some of the potential new treatments for acute abdominal pain? How will technology change the way we diagnose and treat acute abdominal pain?
9. A 25-year-old woman presents to the emergency room with sudden onset of severe abdominal pain. She has also been vomiting and having diarrhea. What do you do?
10. Is it ethical to deny treatment to people with acute abdominal pain who cannot afford it?
11. What are the different tests that can be used to diagnose acute abdominal pain?
12. How does acute abdominal pain compare to other types of pain?

# About the Author

Tom Heston is a Board-Certified family physician and nuclear medicine physician. He brings over three decades of clinical practice, teaching medical students, and engaging in academic writing. While pursuing Music Theory and History at the University of Washington, he explored his passion for music, computer science, and nutrition. Furthermore, he earned a Masters's degree in blockchain technology at the University of Nicosia. Dr. Heston's earned his Medical Doctor degree from St. Louis University School of Medicine, graduating with a Distinction in Research. Throughout his professional journey, he pursued advanced training at renowned institutions such as Duke University, the University of Washington, Oregon Health & Sciences University, and Johns Hopkins University.

As a respected academic, he has previously held teaching positions at Johns Hopkins University and the International American University in St. Lucia. He currently serves as a Clinical Associate Professor at Washington State University and a Clinical Instructor at the University of Washington. He has published over 100 esteemed publications covering medical imaging, family medicine, medical informatics, and biostatics. Early on, during his residency training in the early 1990s, he embarked on pioneering research on using neural networks in medicine, a field he continues to explore alongside biostatistics, blockchain technology, and advances in clinical medicine.

Despite his remarkable medical career, Dr. Heston's artistic inclination remains intact, and he continues to compose and publish music. This holistic approach to life and learning defines his unique perspective. Currently engaged in medical research, he resides with his family in the awe-inspiring Pacific Northwest, drawing inspiration from the region's breathtaking natural beauty.